\documentclass[a4paper,11pt]{article}
\usepackage[utf8]{inputenc}
\usepackage{graphicx}
\usepackage{color}
\usepackage{amsmath}
\usepackage{epstopdf}
\usepackage{verbatim}
\usepackage{wrapfig}
\usepackage{placeins}
\usepackage{tabu}
\usepackage{pdflscape}
\usepackage{wasysym}

\title{Signatures of the Martian rotation parameters \\
in the Doppler and range observables}
\author{Marie Yseboodt, V\'eronique Dehant,  Marie-Julie P\'eters\\
Royal Observatory of Belgium, Avenue circulaire 3, Brussels, Belgium,\\ 
m.yseboodt@oma.be}
\date{Published in Planetary and Space Sciences, 2017}

\begin{document}

\maketitle

\begin{abstract}
The position of a Martian lander is affected by different aspects of Mars' rotational motions:
the nutations, the precession, the length-of-day variations and the polar motion.  
These various motions have a different signature in a Doppler observable between the 
Earth and a lander on Mars' surface.
Knowing the correlations between these signatures and the moments when these signatures 
are not null during one day or on a longer timescale
is important to identify strategies that maximize the geophysical return of observations with a geodesy experiment, 
in particular for the ones on-board the future NASA InSight or ESA-Roscosmos ExoMars2020 missions. 
\\
We provide first-order formulations of the signature of the rotation parameters in the Doppler and range observables.
These expressions are functions of the diurnal rotation of Mars, the lander position,
the planet radius and the rotation parameter. 
Additionally, the nutation signature in the Doppler observable is proportional 
to the Earth declination with respect to Mars.
\\
For a lander on Mars  close to the equator, the motions with the largest signature 
in the Doppler observable are 
due to the length-of-day variations, the precession rate and the rigid nutations.
The polar motion and the liquid core signatures have a much smaller amplitude.
For a lander closer to the pole, the polar motion signature is enhanced while 
the other signatures decrease.
\\
We also numerically evaluate the amplitudes of the rotation parameters signature 
in the Doppler observable for landers on other planets or moons.
\end{abstract}

%\begin{keyword}
%Dynamics; rotation; tracking; interior; Mars; lander
%\end{keyword}

\section{Introduction}
Tracking a lander on the surface of another planet with radio signals is an efficient way to observe 
its rotation (Dehant et al., 2009, 2011).
This can be done by measuring the Doppler shift of the radio signal between the lander and a 
large antenna on Earth, like the ones of the Deep Space Network (DSN).
 Here, we consider a two-way X-band signal: the ground station transmits the radio signal to Mars, 
which is coherently (i.e.\ without any phase shift) sent back to Earth by the transponder on-board 
the lander.
\\

Five Martian landers have already been tracked from Earth in order to 
 determine the Martian rotational parameters.
The first landers were the two Viking landers in 1976-1982
(see for example Borderies et al., 1980, Yoder and Standish, 1997).
Twenty one years later, the Pathfinder lander stayed 3 months on Mars' surface in 1997
(Folkner et al., 1997a and 1997b).
More recently the two Mars Exploration Rovers were tracked from Earth when they were stationary 
(Le Maistre, 2013, Kuchynka et al., 2014): Spirit was stuck at the end of its life in 2009, 
and Opportunity stopped moving during the winter for energy saving during four months in 2012.
The tracking of various orbiters has also largely contributed to the determination of 
Mars rotation (see for example Konopliv et al., 2016).

Two Martian missions including a geodesy experiment will be launched in the coming years.
The first one is the NASA mission InSight (Interior Exploration using Seismic Investigations, 
Geodesy and heat Transport) to be launched in 2018. 
The InSight spacecraft will have a Small Deep Space transponder on board: 
the RISE (Rotation and Interior Structure Experiment, Folkner et al., 2012) transponder.
The second mission is the ESA-Roscosmos mission ExoMars2020 that will include the LaRa (Lander Radioscience, 
Dehant et al., 2009) transponder on the surface platform. This spacecraft is to be launched in 2020. 
\\

The Martian Orientation Parameters (MOP) are the nutations, precession, length-of-day 
variations and polar motion.
The precession is the long-term drift of the rotation axis in space.
The nutations are periodic motions of this axis observed from space,
while the polar motion is the seasonal motion of the rotation axis 
in a frame tied to the planet. 
The rotation rate variations, also called length-of-day 
variations, are the periodic variations of the Martian diurnal rotation.
The periods of these variations correspond to the harmonics of the annual period.
Since these rotational motions depend on the interior structure of Mars, 
in particular the core dimension, density and state, and on the 
dynamics of the atmosphere and the ice caps and the CO$_2$ sublimation/condensation process, 
accurate measurements of these angles are very useful because
additional constraints on the geophysical models could be provided.
\\

Each rotational motion has a different signature in the lander-Earth observable. 
This study aims at describing the behavior of the MOP signatures in different radio observables. 
These expressions are useful in order to understand the role of 
various configuration parameters, like for example the influence of the lander position 
or the configuration geometry and 
to anticipate the observation times that maximize the signatures.
Since parameters are correlated when they are estimated from the same dataset, 
a parameter that has a large signature may have a large uncertainty in its final estimation.
In order to know the expected precision on each geophysical parameter, a full numerical 
simulation with inversion of noisy data has to be perform, which is not the goal of this study.
Nevertheless by knowing the signatures, we can better anticipate the correlations 
between the parameters (analytical expressions of the correlations between
two parameters can be found if their temporal behavior is similar)
and better understand the results of full numerical inversion using different observation strategies. 
This is very relevant for the InSight and ExoMars missions preparation. 
The information from this study can be used to answer rapidly to questions 
related to programmatic, in order to see if a change of the nominal mission parameters 
augment or increase the return of the science case. 
For example, in order to design the lander antenna, it is important to know if there is an 
elevation range that maximizes the MOP signatures.
\\

At the beginning of the space age, Hamilton and Melbourne (1966) analytically expressed in a simple way 
the Doppler observable between a distant spacecraft and the Earth tracking station as a 
function of the equatorial coordinates of the spacecraft (right ascension and declination).
They noticed that the Earth rotation induces a diurnal modulation in the signature. 
Latter Curkendall and McReynolds (1969) extended this analysis using more parameters.
In the Mars-Earth radiolink that we investigate here, we have two terrestrial planets 
with their own rotation rate,
and therefore the Doppler observable includes these two different Hamilton-Melbourne effects
with the two diurnal frequencies.
Since we study how the Martian orientation parameters affect the Doppler
observable, the configuration is reversed here with respect to the situation described in 
Hamilton and Melbourne (1966): the antenna is fixed with respect to the Mars' surface 
while the Earth replaces the distant spacecraft. 
Therefore we mostly see the Martian rotation rate modulation in the MOP signature
and the signatures are functions of the equatorial coordinates of the Earth.
\\

The signatures of the MOP parameters in radio observables have already been investigated in 
previous studies.
The expression of the range observable using Earth based 
coordinates can be found in Estefan et al. (1990).
Le Maistre et al. (2012) gave equations for the LOD signatures in the Doppler observable 
in a simplified configuration, 
Konopliv et al. (2006) gave the signature of a variation of latitude in the Doppler observable and 
Folkner et al. (1997a, equation~1) express the range observable as a function of the 
mission parameters and Mars spin axis position.
Yseboodt et al. (2003) studied the MOP signatures in the orbiter-Earth Doppler link 
in the frame of a geophysical mission including a network of landers on Mars' surface.
Kawano et al. (1999) investigated the MOP signatures in the Same Beam Interferometry 
(SBI) observable for two landers (or more) on Mars' surface, providing some analytical 
expressions of the signatures.
A numerical estimation of the MOP precision using a covariance analysis for Earth-based 
observations of a Martian lander was done by Edwards et al. (1992).
\\
In this study, we explicitly give the first-order expressions of the signatures for all 
the MOP and for the lander position in three different observables (Doppler, range and SBI) and 
we explain the method to derive them. We also evaluate the order of magnitude of these signatures
and use these expressions to investigate the correlations between MOP. 
\\

The paper is organized as follows: 
in section 2, we present the different angles and variables that describe the geometry of 
the Doppler observable.
Section 3 gives a summary of the models for the MOP.
In Section 4, we give the analytical expressions of the signature of these MOP
in the Doppler observable and their numerical values, while 
the section 5 focuses on the signatures in the range observable.
In section 6, we discuss how these signatures are useful for the preparation of 
the future geodesy experiments and investigate the correlations between different parameters.
A description of the MOP signatures in the SBI observable is given in appendix \ref{sec_SBI}.

\section{The mission geometry}

The instantaneous Doppler frequency can be interpreted as the space probe radial velocity 
with respect to the tracking station.
As a first order approximation, it is the projection on the line-of-sight of the difference 
of the velocity vector between the emitter and the receiver. 
It is very sensitive to the configuration geometry, therefore we present in this section 
the different variables that are important for the signatures in the Doppler observable.
\\

\begin{figure}[!htb]
\includegraphics[height=9.cm,width=18cm]{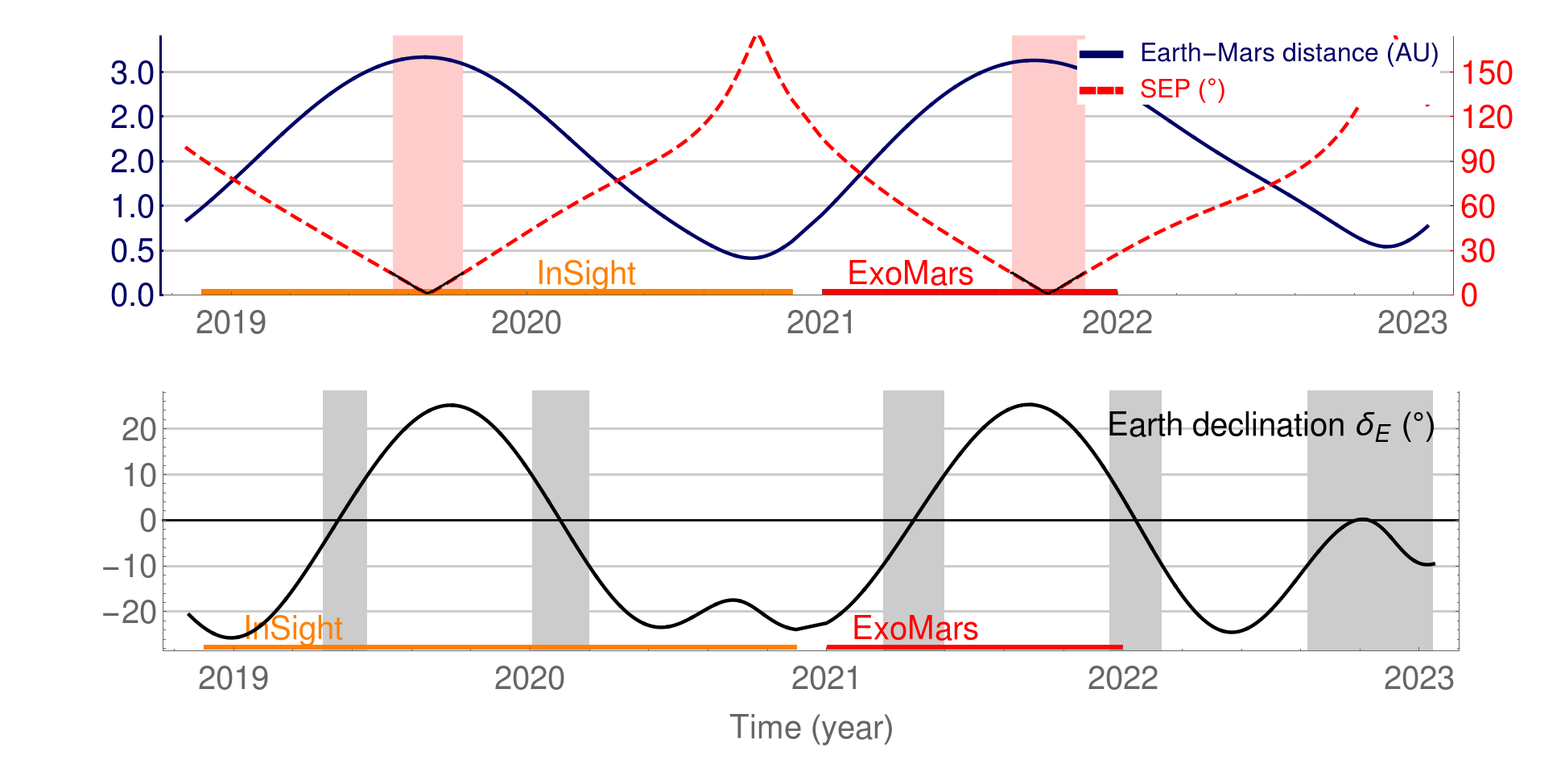}
\caption{Distance between the Earth and Mars and the Earth declination $\delta_E$ with respect to Mars
as a function of time.
The Sun-Earth-Probe (SEP) angle is plotted on the upper graph in red. Units are degrees.
The black parts of the curve (and the corresponding pink boxes) indicate the points where the SEP 
is smaller than $15^\circ$, 
corresponding to radio data with a large noise caused by the solar plasma.
In the lower graph, the gray boxes correspond to time intervals where the Earth declination is
smaller than 10$^\circ$.
The InSight and ExoMars nominal mission lifetime are indicated by the orange and red horizontal lines. 
}
\label{fig_dist}
\end{figure}
The ExoMars and InSight missions have similar Mars-Earth-Sun orbital configuration at launch and 
when arriving on Mars. For both  missions, the distance between Mars and the Earth
increases after the Martian landing, being maximal a few months after the landing 
(see Fig.~\ref{fig_dist}, upper graph).
The Sun-Earth-Probe (SEP) angle or elongation is small when the Earth-Mars distance is maximal, 
this is the conjunction time. 
The pink boxes on Fig.~\ref{fig_dist} indicate where the SEP angle is smaller than $15^\circ$.
At that time, random frequency variations are due to refractive index fluctuations 
along the line of sight caused by phase scintillation as the radio wave passes through
the solar plasma,
and the noise on the radioscience measurements is much larger than the nominal noise.
Noise in one-way propagation at X-band is plotted as function of the SEP on figure 2
of Asmar et al. (2005).
Typical accuracy on two-way X-band Doppler measurements outside conjunction periods is 
about 0.05 mm/s or 0.1 mm/s (2.8 mHz or 5.6 mHz) at an integration time of 60 seconds 
or more (see for example Konopliv et al. 2006).
\\

The Earth declination $\delta_E$ with respect to Mars is the angle between the Mars-Earth direction 
and the equatorial plane of Mars.
Because of the Martian obliquity and the Earth orbital motion in the ecliptic plane, 
the Earth declination has variations caused by the orbital motions of the two planets and by the 
Mars' pole direction. 
It varies between -25$^\circ$ and 25$^\circ$ (see Fig.~\ref{fig_dist}, lower graph).
When the Earth-Mars distance is minimal, corresponding to the opposition, the Earth-Mars direction 
reverses its rotation:
for a few weeks, this direction goes from a prograde to a retrograde motion.
This is caused by the faster orbital velocity of the Earth than of Mars.
At this time, in the Earth's sky, Mars has an apparent retrograde motion with respect to the 
usual planetary rotation 
 and the Earth declination shows some irregularities due to this retrograde motion.
The maxima in the Earth declination do not correspond to the distance minima because the direction of 
the Martian spin axis has to be taken into account as well.
The periodicity of the declination is longer than the Earth-Mars synodic period (780 days). 
On the time scale of a spacecraft mission (about 2 years), the MOP can be safely neglected for the 
computation of the Earth declination,
the largest MOP effect being the Martian precession that changes the Earth declination by less than 
$0.025^\circ$.
The gray boxes on Fig.~\ref{fig_dist} for the Earth declination correspond to time intervals where the 
Earth declination is smaller than 10$^\circ$. In Section 4 we will show that these intervals 
correspond to a lower sensitivity to Mars' spin axis motion in space.

Another useful equatorial coordinate that will appear in the MOP signatures 
(Section 4) is the Earth hour angle $H_E$, the angle along the celestial equator 
from the Mars local meridian 
to the hour circle passing through Earth, in the retrograde direction.
If $\alpha_E$ is the Earth right ascension, $\phi$ the sidereal Mars rotation angle
(the angle from the vernal equinox to the prime meridian along the equator) 
and $\lambda$ the lander longitude (positive east from the prime meridian), 
the Earth hour angle is related to the other angles 
in the equatorial plane by the following equation
\begin{equation}
H_E + \alpha_E = \phi + \lambda
\label{eq_HE}
\end{equation}
This relationship is given in Equation (5) and Figure 15 of Hamilton and Melbourne (1966),
using the following conversion between their notations and ours
$\phi = \alpha_{\astrosun} + \omega \, t_u $, $H_E = \theta - \theta_0$ and $\alpha_E = \alpha$.

The Earth hour angle has a diurnal variation (period of 24h 37min for Mars), 
increasing every day non-linearly from 0 to 360$^\circ$, and depends on the 
lander position. 

The Earth trajectory in the lander sky is quasi identical from one day to the other.
Every day the Earth rises and crosses the lander sky.
Like the Earth declination $\delta_E$, the Earth elevation $h$ has variations related to the orbital 
motion of the Earth and Mars.
If $\theta$ is the lander latitude, the maximal elevation $h_{max}$ of the Earth in the 
lander sky is $90^\circ - | \delta_E - \theta |$.

The RISE instrument has two directional antennas designed with a central axis pointing $28^\circ$ 
above the horizon, 
with one antenna pointing nearly east and the other pointing due west. 
LaRa antennas are omnidirectional in azimuth approximately covering the Earth elevation range between 
$30^\circ$ and $55^\circ$.

\section{The Martian orientation parameters and the motion of the lander}
\label{secMOP}

The rotation matrix from the Mars body-fixed reference frame to the inertial frame
based on Mars Mean Orbit of J2000 is 
(Folkner et al., 1997a, Konopliv et al., 2006, Le Maistre et al., 2012)
\begin{equation}
\mathbf{M} = R_Z(- \psi)\, . \, R_X(- \epsilon) \, .\, R_Z(- \phi) \,.\, R_Y(X_P) \,.\, R_X(Y_P)
\label{eq_M} 
\end{equation}
If $\mathbf{r_{bf}}$ is the lander position vector in the body-fixed reference frame, 
the position of the lander in the inertial frame $\mathbf{R}$ is:
\begin{equation}
\mathbf{R} = \mathbf{M} \, . \, \mathbf{r_{bf}}
\label{eq_R} 
\end{equation}
$R_X$, $R_Y$ and $R_Z$ are the classical elementary rotation matrices around the X, Y, Z axes, respectively.
\begin{equation}
R_X(\theta)=\left ( \begin {array}{ccc} 1&0&0\\\noalign{\medskip}0&\cos \theta&
\sin \theta \\\noalign{\medskip}0&-\sin \theta &\cos \theta
\end {array}\right ) \;
\end{equation}
\begin{equation}
R_Y(\theta)=\left (\begin {array}{ccc} \cos \theta &0&-\sin \theta 
\\\noalign{\medskip}0&1&0\\\noalign{\medskip}\sin \theta &0&\cos 
\theta \end {array}\right ) \;
\end{equation}
\begin{equation}
R_Z(\theta)=\left (\begin {array}{ccc} \cos \theta &\sin \theta &0
\\\noalign{\medskip}-\sin \theta &\cos \theta &0\\\noalign{\medskip}0&0
&1\end {array}\right ).
\end{equation}
As for the Earth, five angles called here the Martian Orientation Parameters (MOP) are used 
to characterize the matrix $\mathbf{M}$: $\epsilon$, $\psi$, $\phi$, $X_P$ and $Y_P$. 
$X_P$ and $Y_P$ are the crust-fixed coordinates of Mars' spin axis related to the polar motion. 
$\phi$ is the Martian sidereal rotation angle. 
Its variations $\delta\phi$ are called the length-of-day (LOD) variations.
$R_X(-\epsilon) $ and $R_Z(-\psi)$ are the Precession-Nutation matrices depending on 
the obliquity $\epsilon$ between the instantaneous Mars equator and 
the mean orbital plane at J2000 and on
the longitude of the node of the Mars equator $\psi$.
Nutation in obliquity and longitude $\delta\epsilon$ and $\delta\psi$ 
are the periodic variations of $\epsilon$ and $\psi$ 
while the precession $\dot \psi$ is the secular trend of $\psi$.
We can write
\begin{eqnarray}
\phi     &=& \phi_{0} + \Omega \,t + \delta\phi   - \delta\psi \cos\epsilon \label{eq_phi}\\
\epsilon &=& \epsilon_0 + \delta\epsilon  \\ 
\psi     &=& \psi_0 + \delta\psi + \dot \psi \, t
\end{eqnarray}
$\Omega$ is the diurnal rotation rate  and $t$ is the time with respect to the J2000 epoch.
The MOP are small angles. 
Tab.~\ref{tab1} gives a summary of the investigated parameters and their amplitudes. 
\\

\begin{table}[!htb]
{\small
\noindent
\begin{tabular}[]{|c|c|l|}
\hline
Parameter &  Frequencies & Amplitudes \\
\hline 
Rigid nutations amplitudes $\delta\epsilon$& $n$, $2n$, $3n$, $4n$, $5n$, $6n$& 49, 515, 113, 19, 3, 0 mas\\
 $\qquad\qquad\qquad\qquad\delta\psi$& $n$, $2n$, $3n$, $4n$, $5n$, $6n$& 555, 1130, 242, 41, 6, 1 mas\\
Amplification from a large liquid core $\delta\epsilon$& $n$, $2n$, $3n$, $4n$, $5n$, $6n$& 6, 12, 9, 1, 0, 0 mas\\
 $\qquad\qquad\qquad\qquad\qquad\qquad\qquad\delta\psi$& $n$, $2n$, $3n$, $4n$, $5n$, $6n$& 4, 39, 3, 1, 0, 0 mas\\
\hline 
Precession rate $\dot\psi$ & & 7576 mas/y \\
\hline 
Rotation rate variations $\delta\phi$ & $n$, $2n$, $3n$, $4n$ & 505, 139, 35, 13 mas \\
\hline 
Polar Motion $X_P$ & $n$, $2n$, $4n$, $\omega_{Ch}$ & 13, 8, 2, 55 mas\\
$\qquad\qquad \qquad Y_P$ & $n$, $2n$, $4n$, $\omega_{Ch}$ & 5, 8, 2, 34 mas\\
\hline
\end{tabular}
\caption{The Mars rotation parameters (MOP) that are investigated.
The explanation of the models and references are given in the text.}
\label{tab1}
}
\end{table}

Nutation is the periodic motion of the spin pole in space.  
The nutations for a rigid planet can be computed from celestial mechanics and ephemerides.
A rigid nutation model is usually constructed as a sum of multiple frequencies with an amplitude 
and a phase for the nutations in obliquity ($\delta\epsilon$) and in longitude ($\delta\psi$),
 see Eq.~(\ref{equationnut}) for $\delta\epsilon$, equation for $\delta\psi$ being similar.
\begin{eqnarray}
\delta\epsilon(t) &=& \sum_{i=1}^6{\left(\delta\epsilon_{i}^S \sin(i\, n\, t) + \delta\epsilon_{i}^C \cos(i\, n\, t) \right)} 
\label{equationnut}
\end{eqnarray}
where $n$ is the annual frequency of Mars.
In the following sections, we use the term ``nutation in longitude'' $\delta\psi$ to mean the sum 
of the components at
multiple frequencies of the nutation in longitude, and similarly, 
the term ``nutation in obliquity'' $\delta\epsilon$, as the sum of the main nutations 
at multiple frequencies.
For Mars, the largest nutations are the annual and semiannual nutations. 
Their maximal amplitude are about 555 and 1130 mas, corresponding to 9 and 19 m on Mars surface. 
We use Roosbeek (1999) rigid Mars nutations model. 
\\
Another formulation can also be used: prograde $p$ and retrograde $r$ nutation rather than 
nutations in obliquity $\delta\epsilon$ and in longitude $\delta\psi$.
The following equation links the two formulations for each frequency $i$:
\begin{equation}
p_i \, e^{i \, n \, I \, t} + r_i \, e^{- i \, n \, I \, t} = 
\delta\epsilon_{i}^S \sin(i\, n\, t) + \delta\epsilon_{i}^C \cos(i\, n\, t) 
+ I \sin \epsilon \left(\delta\psi_{i}^S \sin(i\, n\, t) + \delta\psi_{i}^C \cos(i\, n\, t) \,
\right)
\end{equation}
where $I=\sqrt{-1}$.

However Mars is not a rigid planet. Therefore the nutation amplitudes depend 
on the interior structure and parameters, like the state (liquid or solid) of the core or its 
moment of inertia.
We can model the non-rigid nutation amplitudes by applying a transfer function to the rigid nutations
(Dehant et al., 2000, Van Hoolst et al., 2000a, 2000b). 
This transfer function is based on the model developed for the Earth by Sasao et al.~(1980) 
and includes the liquid core effects and other deformation effects.
If the core is liquid, then an additional free mode, the free core nutation (FCN)
amplifies the nutation response, for example the prograde annual nutation.
Current interior models estimate the FCN period to be between about -240 and -285 days.
Therefore the FCN can also enhance the small retrograde terannual (229 days) or the large prograde 
semiannual nutation (343.5 days).  
Considering the 3 sigma interval rather than the 1 sigma interval makes the FCN range larger than 
$(-240,-285)$ days 
and the FCN frequency can be even closer to the terannual frequency, increasing the amplification.
In this study, we use a transfer function and amplifications derived from up-to-date interior models 
(Panning et al., 2016). The mantle mineralogy and the temperature profile are unique 
for all the models and the planet is assumed to be in hydrostatic equilibrium.
The variable quantities in the interior models are the size of the core and the density and thickness of the crust.
The models agree with the observed moment of inertia at one sigma.
\\

The frequencies of the nutations are not affected by the transfer function, therefore we use a 
trigonometric expansion with known periods but unknown amplitudes (Eq.~(\ref{equationnut})). 
The amplifications are different for each nutation period and 
for the nutation in obliquity and in longitude 
(see for example Figure 10 of Le Maistre et al., 2012).
Depending on the size of the core, the maximal change of the nutations caused by the liquid core 
is between $11$~mas and $24$~mas (corresponding to about $18$~cm and $40$~cm on Mars surface) for the nutation in obliquity
and between $24$~mas and $40$~mas or $39$~cm and $66$~cm for the nutation in longitude.
The lower values correspond to a small core of $1 400$~km while the upper values correspond to a 
large core of about $1 900$~km.
In this study, we present results for 2 models among all the interior models: 
one with a large core (with a core-mantle boundary (CMB) radius of 1857 km and a FCN period of -242 days) 
and one with a small core (CMB radius of 1417 km, FCN period of -284 days).
An important objective of the geodesy experiments is to measure the nutations and particularly the 
transfer function since it is a good tool to constraint the interior structure of Mars and the core 
dimension and density in particular.
\\

The precession is the secular motion of the spin axis about the normal to the orbit plane. 
The spin axis completes one rotation in a period of about 170 340 years. 
The precession is caused by the gravitational pull of the Sun on the oblate figure and depends 
on the moment of inertia about the spin axis. 
The present uncertainty on the precession rate (Konopliv et al., 2016) is 2.1 mas/y,
corresponding to a period change of about 47 years.
Since the precession rate $\dot \psi$ is proportional to $1/C$ where $C$ is Mars moment of inertia,
the estimation of the precession rate gives a constraint on the interior mass distribution.
\\

The rotation rate variations or length-of-day (LOD) variations $\delta\phi$ are modeled by a 
trigonometric series with the annual, semiannual, terannual and quaterannual periodicities, 
see Eq.~(\ref{equationUT}).
\begin{eqnarray}
\delta\phi(t) &=& \sum_{i=1}^4{\left(\delta\phi_{i}^S \sin(i\, n\, t) + \delta\phi_{i}^C \cos(i\, n\, t) \right)} 
\label{equationUT}
\end{eqnarray}
The annual nutation has an amplitude of about $505$~mas or $8$~m on Mars surface 
while the semiannual nutation is about $139$~mas or $2.2$~m (Konopliv et al., 2016).
In the rotation matrix around the Z-axis ($\phi$ angle, see Eq.~(\ref{eq_phi})), 
above the LOD variations, there is an additional term depending 
on the nutation $- \delta\psi \cos \epsilon$, adding a nutation signature in the Doppler observable.
The LOD variations are of interest for global climate models, to constrain the atmosphere and ice caps dynamics.
\\

Each polar motion (PM) coordinate $X_P$ and $Y_P$ is modeled by a trigonometric series 
(annual, semiannual, quaterannual) and a Chandler wobble term,  see Eq.~(\ref{equationPM})
for $X_P$, the equation for $Y_P$ being similar.
\begin{eqnarray}
X_P(t) &=&  X_{Ch}^S \sin(\omega_{Ch} \,t) + X_{Ch}^C \cos(\omega_{Ch}\, t)  + 
\sum_{i={1,2,4}}{\left( X_{i}^S \sin(i\, n\, t) + X_{i}^C \cos(i\, n\, t) \right)} 
\label{equationPM}
\end{eqnarray}
where  $\omega_{Ch}$ is the Chandler frequency.\\
The amplitudes of the seasonal terms of the polar motion are expected to be between 0 and 15 mas
(Defraigne et al., 2000, Van den Acker et al., 2002).
The period of the Chandler wobble for Mars is expected to be around 200 or 220 days.
Its amplitude has not yet been measured but is expected to be somewhere between 10 and 100 mas. 
The polar motion has both climatological and geodetic information; 
particularly the Chandler wobble period depends mainly on the dynamical flattening of
the planet and it provides information on the planet’s elasticity and
on inelastic behavior.

\begin{table}[!htb]
\begin{tabu}{|l|c|l|c|l|}
\hline
\rowfont[c]{\bfseries} \multicolumn{2}{|c|}{MOP}  & \multicolumn{2}{c|}{Change of the lander position} \\
\rowfont[c]{\bfseries}     & max value  &  Analytical formulation  & max value \\
\hline
Rigid nutations in long. $\delta\psi$ & 29.4 m & 
$R \,\delta\psi\,\sin \epsilon \sqrt{1-\cos^2 \theta \sin^2 (\phi+\lambda)}$ &
12.3 m \\
Rigid nutations in obliq. $\delta\epsilon$& 10.3 m & 
$R \,\delta\epsilon \sqrt{1-\cos^2 \theta \cos^2 (\phi+\lambda)}$ & 10.3 m \\
\hline
\multicolumn{1}{|l|}{Large liquid core} & 0.4/0.7 m & & 0.4 m \\  
\multicolumn{1}{|l|}{Small liquid core} & 0.2/0.4 m & & 0.2 m \\
\hline
Precession: & 0.8 m & 
$R \,\dot\psi \, t \, [ \cos\theta ^2 \, \cos(\phi + \lambda)^2 $ & 0.8 m \\
(if $\Delta\dot\psi =$ 2.1 mas/y)& & 
$ +  \, (\sin\epsilon \sin\theta - \cos\theta \cos\epsilon \sin(\phi + \lambda)\,)^2]^{1/2}$ & \\
\hline
Rotation rate variations $\qquad \delta\phi$ & 11.8 m & $R \,\delta\phi \, \cos \theta $ & 11.8 m \\
\hline
Polar motion $\;$ ($X_P, Y_P$)& 1.2 m & $R \, X_P \, \sqrt{\sin^2\theta + \cos^2\theta \, \cos^2\lambda}$ & 1.2 m \\
(if Chandler amp. $= 1.2$ m)  &       & $+  \, R \, Y_P \, \sqrt{\sin^2\theta + \cos^2\theta \,\sin^2\lambda}$ &     \\
\hline
\end{tabu}
\caption{Maximal values of the MOP and norm of the lander displacement caused by the MOP.
$\theta$ is the lander latitude, $\lambda$ is the lander longitude, $R$ is Mars radius, 
$\phi$ is Mars diurnal rotation angle, $\epsilon$ is Mars obliquity.
}
\label{tab2}
\end{table}

The maximal value of each MOP on a long time interval is given in the second column of Tab.~\ref{tab2}, 
assuming the nominal model. The amplitude of the Chandler wobble is assumed to be around $1.2$ m.
For the liquid core effect in the nutations, the first numbers (0.4 m for a large core and 0.2 m for a 
small core) are the components of the nutation in obliquity, while the second (0.7 and 0.4 m) are 
the components of the nutation in longitude. The precession rate is very large, we consider here only the 
effect of the present uncertainty on the precession rate at J2000 (about 2.1 mas/y).
The displacement produced by this uncertainty after about 20 years after J2000 is about 0.8 m. 

Each MOP changes of the lander position.
If the position change vector is the difference between the lander position 
with and without the MOP and using the rotation matrix $\mathbf{M}$ (see Eq.~(\ref{eq_R})), 
we derive the norm of this position change vector. 
The analytical formulations and the maximal value of the norm of this position change vector 
is given in the last columns of Tab.~\ref{tab2}.
Since we use first order approximation, all the displacements are proportional to the MOP.
The validity of this assumption has been numerically checked, 
second order terms are negligible.
Some of the lander displacements are modulated by the diurnal rotation.
The lander displacement caused by the LOD variations is larger 
when the lander is closer to the equator.
The maximal value of the displacement caused by the rigid nutation is independent of the lander position 
(because the square roots in Tab.~\ref{tab2} are always equal or smaller than $1$), 
but close to the pole, the displacement caused by 
the nutations does not have any diurnal variations. 
There is a reduction of a factor 2.3 from the maximal value of the nutation in longitude $\delta\psi$ to 
its change in the lander motion (from $29$ m to $12$ m)
because the obliquity of Mars ($\epsilon = 25^\circ$) induces a projection in the lander displacement
and because the rotation angle component (- $\delta\psi \,\cos\epsilon$) removes a part of the signal
(see Eq.~(\ref{eq_phi})).
For the other MOP, the maximal displacement caused by the MOP is the same as the MOP maximal value. 
Depending on the size of the core, the maximal change of the lander position due to the liquid core is
between 19 cm (small core) and 42 cm (large core). 
The analytical expression is a combination of the transfer function and the change of lander position
caused by the rigid nutations in obliquity and in longitude, its expression is lengthy and 
therefore has not been included in Tab.~\ref{tab2}.
The nutations in longitude and in obliquity together changes the lander by about 12.6 m.

\section{MOP signature in the Doppler observable}

In this section we give the analytical expressions of the MOP signature in the Doppler 
lander-Earth observable (= direct-to-Earth or DTE observable).
We consider an instantaneous Doppler observable (with an infinite speed of light 
and no integration time).
The Doppler measurements are highly sensitive to the position variations along the line-of-sight
and not sensitive to position variations in the plane perpendicular to the line-of-sight.
\\

The Doppler measurements between the Earth and a spacecraft on Mars varies between about 
-17 and 17 km/s.
This corresponds to the value of the relative velocity of the two planets around the Sun.
Its main modulations come from the relative orbital motions of the Earth and Mars.
Other modulations come from the variation of the orientation and the rotation of Mars and the Earth.
The diurnal rotations of Mars and of Earth impress a sinusoidal modulation upon
the Doppler signal, with a beat period of about 36 days.
This period of 36 days corresponds to the Mars-Earth synodic period of rotation given 
that the rotation period of Earth and Mars are very close, 23h56min versus 24h37min 
(Le Maistre 2013).
\\

The Doppler observable $q$ can be modeled as the projection of the velocity difference $\vec{\Delta V}$ 
on the line-of-sight $q = |\vec{\Delta V}|\,\cos \alpha$, with $\alpha$ being the angle between 
$\vec{\Delta V}$ and the Earth-lander line-of-sight.
The signature of a parameter in the Doppler observable is the observable estimated
taking this parameter into account minus the observable without considering it.
To the first order in the MOP, it can be expressed as (Yseboodt et al., 2003):
\begin{equation}
\Delta q_{MOP}  \approx  \Delta \alpha_{MOP} |\vec{\Delta V}|\, \sin \alpha + 
|\vec{\Delta V}_{MOP}|\, \cos \alpha + \cdots
\label{eq_signatu}
\end{equation}
where $|\vec{\Delta V}_{MOP}|$ is the change in $|\vec{\Delta V}|$ caused by the MOP
and $\Delta \alpha_{MOP}$ is the change in the line-of-sight direction caused by the MOP.
The first term in Eq.~(\ref{eq_signatu}) is the ``geometrical effect'' or 
change of the line-of-sight direction. 
This term expresses the change in direction of the emitter-receiver line induced by the rotation parameter
multiplied by the amplitude of the velocity. 
The second term of Eq.~(\ref{eq_signatu}) is the contribution 
of the velocity change induced by the rotation.
Both terms are proportional to Mars radius and to the MOP.
The line-of-sight displacement due to the MOP $\Delta \alpha_{MOP}$ decreases if the 
emitter-receiver distance increases, therefore the geometrical effect is proportional to the 
relative velocity over the distance $|\vec{\Delta V}|/\vec R_{EM}$. 
The velocity effect, i.e.\ the change of the lander velocity $|\vec{\Delta V}_{MOP}|$, 
is independent of the emitter-receiver distance but is proportional to the diurnal rotation rate $\Omega$.
If the relative velocity over the distance 
is smaller than the planet diurnal rotation rate, then the velocity effect dominates the signature 
in the Doppler observable. This happens when the Earth-planet distance is large. 
Since this is the case for Mars and since Mars diurnal rotation rate $\Omega$ is large, 
this velocity effect is the one that is modeled in this section
while the line-of-sight displacement is neglected.
If the emitter-receiver distance is small, like for example if the receiver is on board an orbiter, 
then the dominant effect in the Doppler signal is the change in the line-of-sight direction.
\\

\begin{figure}[!htb]
\hspace*{-.5cm}
\includegraphics[height=18cm,width=18cm]{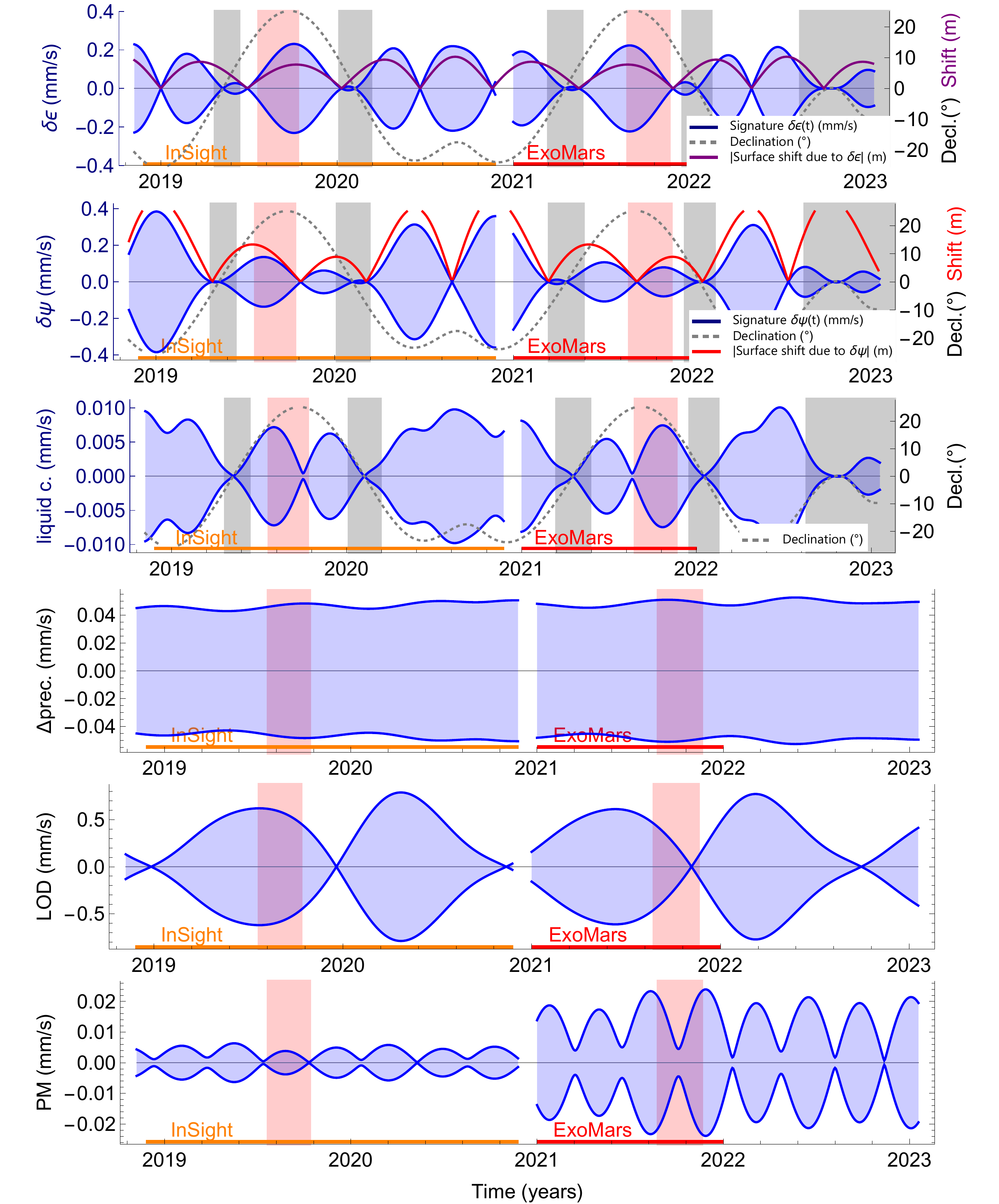} 
\caption{Temporal evolution of the MOP signature in the Doppler observable (in mm/s) 
from top to bottom: the (rigid) nutations in obliquity $\delta\epsilon$, 
the nutations in longitude $\delta\psi$,
the liquid core effect in the nutations (large core, Free Core Nutation period of 242 days), 
a precession rate $\dot \psi$ difference of 2 mas/y, the LOD variations and the polar motion. 
On the left part of the graph, the InSight lander is in Western Elysium Planitia 
($\theta = 4^\circ$N, $\lambda = 136^\circ$E)
and on the right part, the ExoMars lander is in Oxia Planum ($\theta = 18.2^\circ$N, $\lambda = 335.45^\circ$E).
In the nutation plots, the dotted gray line indicates the Earth declination $\delta_E$.
The gray boxes correspond to time intervals where the Earth declination is
smaller than $10^\circ$ while the pink boxes indicates where the SEP angle is  
smaller than $15^\circ$.
The red and purple curves are the surface displacements caused by the
nutation in longitude and nutation in obliquity, respectively.
}
\label{fig_sensi}
\end{figure}
The signatures of the different MOP in the Doppler observable are plotted on Fig.~\ref{fig_sensi}.
Since we use first order approximation, the MOP signatures are proportional to the MOP themselves.
Therefore the MOP periodicities (annual, semiannual, Chandler period...) can also be seen in 
the signatures plot.
The expected numerical values are summarized in section \ref{secamp}.

\subsection{The signature of the nutation}
\label{secnut}
If $\delta\epsilon$ is the nutation in obliquity and $\delta\psi$ the nutation in longitude,
their signature in the lander-Earth Doppler observable is 
\begin{eqnarray}
\Delta q_{\delta\epsilon} & = & - \sin \delta_E \, \Omega\, R\,\cos \theta \cos(\phi+\lambda) 
\,\delta\epsilon
\label{eq_signeps}
\\
\Delta q_{\delta\psi} & = & - \sin \delta_E \, \Omega\, R\,\cos \theta \sin(\phi+\lambda) 
\, \delta\psi\, \sin \epsilon
\label{eq_signpsi}
\end{eqnarray}
where $R$ is Mars radius,
$\Omega$ the diurnal rotation rate of Mars,
$\phi$ the diurnal rotation angle of Mars,
$\theta$ the lander latitude, 
$\lambda$ the lander longitude,
$\delta_E$ the Earth declination with respect to Mars, 
$\epsilon$ Mars' obliquity,
$H_E$ the Earth hour angle and 
$\alpha_E$ the right ascension.
The mathematical derivation of these expressions is given in appendix \ref{ap1}.
\\
We can also express the signature as a function of the prograde and retrograde nutations.
\begin{equation}
\Delta q_{nut_{i}} = - \sin \delta_E \, \Omega\, R\,\cos \theta \left(  
p_i^A\cos(i \, n \, t - \lambda - \phi + \varphi_i^p) + 
r_i^A\cos(i \, n \, t + \lambda + \phi - \varphi_i^r) \right)
\label{eq_retro}
\end{equation}
where $\varphi_i^p$ and $\varphi_i^r$ are the phase of the prograde and retrograde nutations.
\\

Since the signature is proportional to the nutation, the annual, semiannual and terannual modulations 
in the $\delta\epsilon$ and $\delta\psi$ nutations also appear 
in the plots, see for example the blue, red and purple lines in Fig.~\ref{fig_sensi}.
The time derivatives of the nutations $\delta\psi'$ and $\delta\epsilon'$ give second order effects in the signature 
since their period (mostly annual and semiannual) is much larger than the diurnal rotation rate $\Omega$.
The nutation signature is proportional to the distance between the lander and the spin axis 
$R \cos \theta$, therefore 
the nutation signature is maximal at the equator and null at the pole.
The nutation signature in the Doppler observable 
must be a function of the direction of its spin axis with respect to the Earth direction,
Eqs.~(\ref{eq_signeps}) and (\ref{eq_signpsi}) show that it is proportional to the sinus of the Earth declination.
A tilt of the rotation axis due to the nutation tilts the diurnal 
velocity vector and the resulting velocity difference is, to the first order, along the Z (polar) axis 
in a body-fixed reference frame.
The projection of this velocity vector on the line-of-sight is null if the Earth in the equatorial 
plane of Mars ($\delta_E=0$). 
Therefore the Doppler observable sensitivity to nutations is low when Mars' spin axis is nearly perpendicular 
to the Mars-Earth direction (see the gray line and gray boxes in Fig.~\ref{fig_sensi}).
Replacing $\sin\delta_E$ by $\delta_E$ leads to a maximal error of 3 \% because the Earth declination 
is always smaller than Mars' obliquity ($-25.2^\circ \le \delta_E \le 25.2^\circ$). 
The factor $\sin\delta_E$ in the nutation signature varies between $-0.42$ and $0.42$, therefore 
decreasing the signatures.
In Eqs.~(\ref{eq_signeps}) and (\ref{eq_signpsi}), there is a diurnal periodicity in the signature 
through Mars' rotation angle $\phi$ or through the local Earth hour angle $H_E$. 
The link between the angles in the equatorial plane is given in Eq.~(\ref{eq_HE}).
This diurnal variation is indicated on the plot by the blue regions between the maximal and minimal curve.
During one day, the signature of $\delta\psi$ and $\delta\epsilon$ cannot be maximal at the same time, 
there is a time shift of 6 hours between the maxima of the two nutations.
\\
The signature of the prograde and retrograde nutations also has a diurnal variation.
However, its frequency is not strictly $\Omega$ but is slightly changed by the harmonics of the 
annual period (see Eq.~\ref{eq_retro}). 
If plotted on a long time scale, for example 700 days, the annual modulation is not visible. 

The Earth elevation in the lander's sky does not appear in the nutation signature.
However there is an indirect link between both quantities, 
since the Earth elevation depends on the chosen schedule of observation.
For example, if observations are chosen every early morning, the Earth elevation is always small,
and from day to day, the value of the Earth hour angle is almost the same.

\subsection{The signature of the liquid core}
\begin{figure}[!htb]
\hspace*{-2.cm}
\includegraphics[height=9cm,width=20cm]{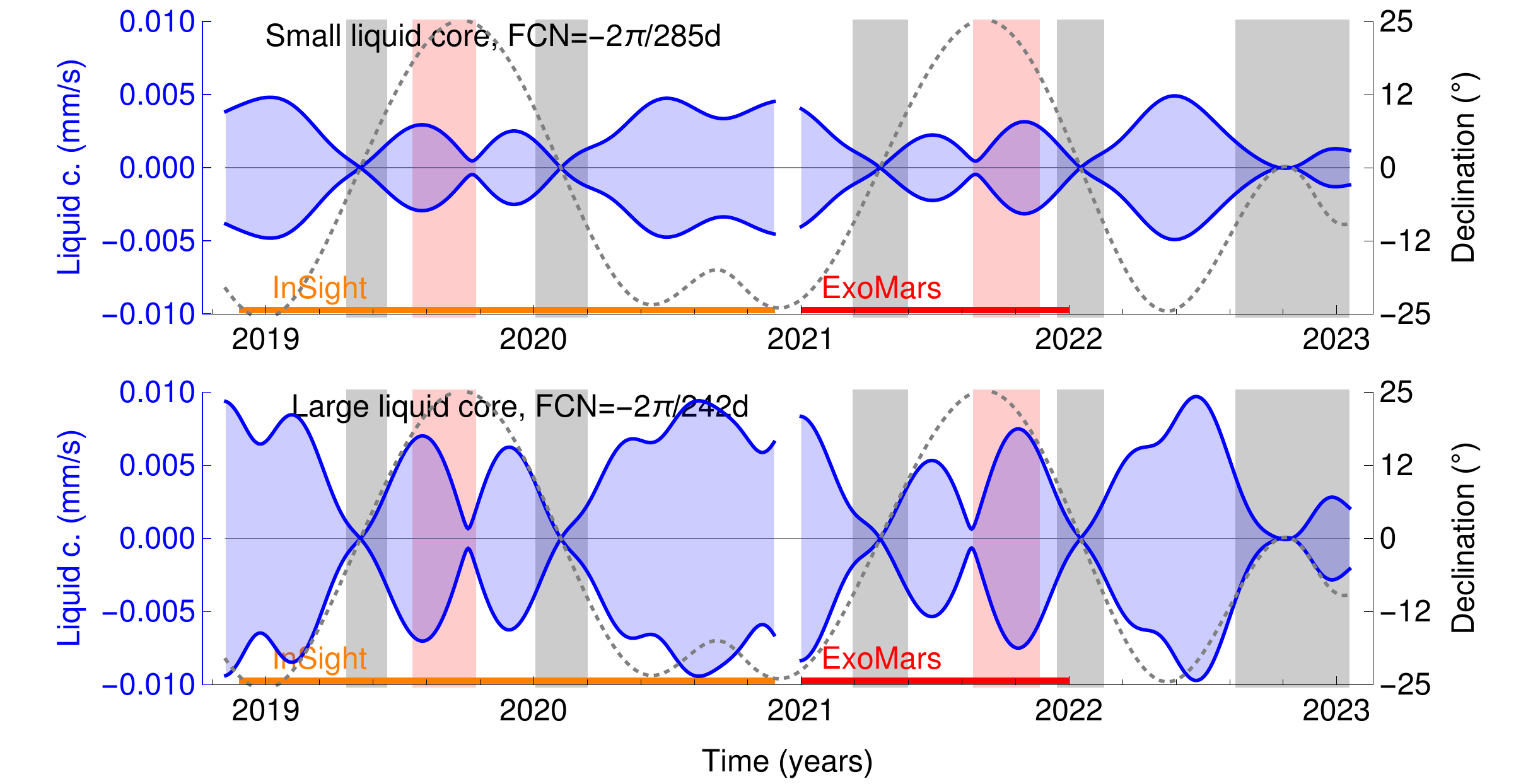} 
\caption{Temporal evolution of the liquid core signature in the Doppler observable (in mm/s)
for two different interior models: a large core and a small core. 
The dotted gray line is the Earth declination $\delta_E$.
}
\label{fig_liquid}
\end{figure}

The liquid core signature in the Doppler observable is the difference between the Doppler observable 
with and without the nutation transfer function, each part being the sum of the $\delta\epsilon$ and 
$\delta\psi$ contributions (Eqs.~(\ref{eq_signeps}) and (\ref{eq_signpsi})).
The liquid core signature is much smaller than the rigid nutations signature
because the liquid core amplification is also smaller. 
The signature of the non-rigid part of the nutations 
has more modulations than the other MOP signatures because sometimes
there is an amplification of the nutation in longitude caused by the liquid core 
while later, it is the nutation in obliquity that is amplified. 
Therefore, the liquid core signature is a combination of both signatures. 
If plotted on the same figure, the rigid and non-rigid nutation signatures are very close and 
indistinguishable on Fig.~2.
The dependence of the liquid core signature on the Earth declination 
(given by the gray line and gray boxes) and 
the lander latitude is the same as that of the rigid nutation signature.
Over an interval of a few years, the position of the maxima and of the roots moves if 
the Free Core Nutation (FCN) period changes (see Fig.~\ref{fig_liquid}).
This is because, if the FCN period is close to the terannual period ($-229$ days), 
the terannual nutation will be largely amplified, while, if the FCN period is close 
to the semiannual period ($-343.5$ days), the terannual retrograde nutation 
amplification will be small but the semiannual nutation will be larger, 
see for example the figures 3 and 4 in Dehant et al. (2000) or equation 7 in 
Folkner et al. (1997a).
\\

Again, the diurnal variation is shown on the plot by the blue area.
Therefore depending on the lander position, each day there is a different observation time
that maximizes the liquid core signature. 
Over 12 hours, the liquid core signature changes from a maximum to a minimum.
6 hours later, the liquid core signature is null.
A large negative signature gives a geophysical information as important as a large positive signature.
The time where the liquid core signature is maximal during the day depends on the FCN value.
The time difference is smaller than an hour for the FCN range investigated here.

\subsection{The signature of the precession}
\label{secprec}
Using a method similar to the one used in appendix \ref{ap1}, we find the expression of 
the precession rate $\dot\psi$ signature in the Doppler observable.
If the product $\dot\psi \,t$ is assumed to be small, the precession signature is 
\begin{equation}
\Delta q_{\dot\psi} = \Omega \, R \, \dot\psi \, t \, \cos\theta \,( 
\cos \delta_E \, \cos H_E  \, \cos\epsilon 
- \sin \delta_E \, \sin(H_E + \alpha_E)\, \sin \epsilon \,)
\label{eq_signpsip} 
\end{equation}
The signature increases with time but remains relatively small on an interval like the mission duration, 
which is short with respect to the precession period of about $170\,000$ years. 
The precession signature has a diurnal variation via the Earth hour angle $H_E$.
\\
The first term in Eq.~(\ref{eq_signpsip}) is the dominant one close to the equator because 
$\cos \delta_E \, \cos\theta \, \cos\epsilon\approx 0.9$. 
It depends on $\cos\epsilon$ while the second term depends on $\sin \epsilon$. 

The maximal signature is almost constant with time (see Fig.~\ref{fig_sensi}), 
there is a long period variation with a small amplitude due to $\cos \delta_E$ 
in the first term of Eq.~(\ref{eq_signpsip}) and due to $\sin \delta_E$ in the second term.

The second term (but not the first one) appears in the signature of the nutation in longitude 
$\delta\psi$.
There is a large difference for the analytical expressions of $\delta\psi$ and $\dot\psi \, t$ 
since the nutation in longitude appears twice in the rotation matrix $\mathbf{M}$
(in $R_Z(- \psi)$ and in Mars' rotation angle rotation $R_Z(- \phi)$)
while the precession appears once in $R_Z(- \psi)$.

\subsection{The signature of the LOD variations}
\label{secLOD}
The signature of Martian rotation angle variations $\delta\phi$ (or LOD variations) 
in the Doppler observable is
\begin{equation}
\Delta q_{\textrm{LOD}} = \Omega\,R \, \cos \theta \cos H_E \, \cos\delta_E \, \delta\phi
\label{eq_UTsign}
\end{equation}
There is a proportionality between the LOD variations and its signature, 
therefore the main frequencies are annual, semiannual, terannual and quaterannual.
Similarly as for the nutation, the LOD signature is proportional to the distance from the lander to the spin axis 
$R \cos \theta$, therefore the signature is maximal at the equator and null at the pole.
The lander longitude disappears from the signature. However there is an indirect effect since 
the Earth hour angle $H_E$ depends on the longitude.
The signature has again a diurnal periodicity via the Earth hour angle. 
The LOD signature is very little affected by the variations of the Earth declination since the 
$\cos\delta_E$ stays between $0.9$ and $1$.
\\

The diurnal velocity vector is in the equatorial plane of the body-fixed reference frame and is perpendicular 
to the lander position vector $\mathbf{r}_{bf}$.
The change of the velocity vector caused by the LOD variations is also in the equatorial plane and 
to the first order perpendicular to the diurnal velocity, it is in the 
direction of the lander position vector $- \mathbf{r}_{bf}$. 
The scalar product between the velocity difference and the line-of-sight (LOS) is
maximized when the two vectors are in the same direction, therefore when the LOS 
is ``aligned'' with the Mars-lander vector. 
Each day, the LOD signature is maximized when the Mars-Earth direction crosses 
the local meridian of the lander (when the local Earth hour angle is null) or equivalently, 
when the Earth culminates in the lander sky. 
Typically this happens near the local noon.
When the lander is on the equator ($\theta = 0$), the maximal Earth elevation is 
$h_{max} = 90^\circ - |\delta_E|$ (see section 2). 
Therefore the maximal LOD signature in the Doppler observable is proportional to $\sin h_{max}$ and 
on a long time interval, the signature increases as the maximal Earth elevation increases 
or equivalently, as the Earth declination decreases to $0$.
When the lander is not on the equator, the maximal LOD signature in the Doppler observable
is again when the local Earth hour angle is null, corresponding to the maximal Earth elevation 
during that day.
However the LOD signature is no longer proportional to $\sin h_{max}$.
Eq.~(\ref{eq_UTsign}) can be rewritten as a function of $h_{max}$:
\begin{equation}
\Delta q_{\textrm{LOD}} = \Omega\,R \, \cos \theta \sin(h_{max}+\theta) \, \cos H_E \, \delta\phi \qquad.
\label{eq_UTsign2}
\end{equation}

\subsection{The signature of the lander position}
\label{secXY}
The signature in the Doppler observable of the equatorial plane coordinates of the lander $X$ and $Y$ 
in the body-fixed reference frame is
\begin{equation}
\Delta q_{\textrm{XY}} = \Omega \, \cos\delta_E \, (X \, \sin(H_E-\lambda) + Y \, \cos(H_E-\lambda)) 
\label{eq_XYsign}
\end{equation}
There is always a diurnal periodicity.
On a daily basis, the $X$ signature is maximized when $H_E-\lambda = \pi/2$ or $3 \pi/2$ while 
the $Y$ signature is maximized when $H_E-\lambda = 0$ or $\pi$.

If we express this signature as a function of the lander latitude $\theta$, the lander longitude disappears 
from the equation
\begin{equation}
\Delta q_{\textrm{XY}} = \Omega \, R\, \cos \theta \, \cos\delta_E \, \sin H_E 
\label{eq_XYsignlat}
\end{equation}

It is known that the Doppler measurements are not very sensitive to the $Z$-coordinate of the lander
(along the polar axis), and therefore solving for this coordinate is difficult (Le Maistre, 2016).
A change in the $Z$-coordinate only does not change the lander diurnal velocity.
Therefore its signature is very small because this is a second order contribution:
the direction of the line-of-sight is changed a little bit due to the lander displacement 
in the polar direction.
There is also a small second order velocity change.
However if we want to know the third component of the lander position, we can impose that 
the lander has to stay on Mars' surface and there is a relationship between the three coordinates. 
This additional constraint related to the topography model helps to solve for the $Z$-coordinate 
(Le Maistre, 2016).

Using the same equations, we can also evaluate the signature of a tidal displacement.
A lander displacement along the X and Y axis caused by the tides has a signature in the
Doppler observable given by Eq.~(\ref{eq_XYsign}).

\subsection{The signature of the polar motion}
\label{secPM}
The signature of the polar motion $X_P$ and $Y_P$ in the Doppler observable is
\begin{equation}
\Delta q_{\textrm{PM}} = \Omega \, R \, \sin \theta \, \cos\delta_E \,
(- X_P \, \sin(H_E-\lambda) + Y_P \, \cos(H_E-\lambda)) 
\label{eq_PMsign}
\end{equation}
To the first order in the polar motion parameters, the first two coordinates of the position 
of the spin pole are $(X_P, - Y_P)$, explaining the sign difference when comparing 
this equation to Eq.~(\ref{eq_XYsign}).
The diurnal periodicity in the signal comes from the Earth hour angle $H_E$.
As for the LOD variations, the signature is not very sensitive to the Earth declination.
During one day, the maxima of the $X_P$ signatures and the minima of $Y_P$ signatures happen at different time, 
separated by 6 hours.
The polar motion signature is a function of $\sin \theta$, therefore the signature increases with 
the lander latitude, being maximal at the poles.
The reason is that the shift of the lander position caused by the polar motion is
$ R \,(X_P \sin\theta,\, - Y_P \sin\theta,\, \cos\theta (Y_P \sin\lambda - X_P \cos\lambda) )$.
Since the diurnal velocity vector is in the equatorial plane, because of the scalar product of 
the Doppler observable, the part of the lander displacement that is important is also in the equatorial plane. 
This lander displacement in the equatorial plane is proportional to $\sin\theta$.
Equivalently to the sensitivity to the lander $Z$-coordinate, the Doppler observable is not sensitive 
to the $Z$ displacement of the lander caused by the polar motion.

\subsection{Numerical evaluation of the signatures for a Martian lander}
\label{secamp}
The maximal value of the MOP signature in the Doppler observable are given using
the InSight and ExoMars nominal missions configuration in Tab.~\ref{tablsi}. 
The nominal mission time coverage is taken into account: 2 Earth years for the InSight 
lander starting end of 2018, 1 Earth year for the ExoMars lander starting beginning of 2021. 
The InSight landing site is expected to be close to the equator in Western Elysium Planitia 
($\theta = 4^\circ$N, $\lambda = 136^\circ$E) 
while we pick the Oxia Planum location ($\theta = 18.2^\circ$N, $\lambda = 335.45^\circ$E)
for the ExoMars lander.
The temporal evolution of the MOP signature in the Doppler observable 
is plotted on Fig.~\ref{fig_sensi}. 
Except for the polar motion signature, 
moving the lander position from one landing site to the other
changes the signatures by the factor $\cos 18.2^\circ/\cos 4^\circ = 0.95 \approx 1$,
which is barely noticeable on Fig.~2 (see the left and the right parts).

\begin{table}[!htb]
\centering
 $\begin{array}{|c|c|c|}
\hline
\text{MOP signature } & \text{InSight} & \text{ExoMars} \\
\hline
\text{Nutations in obliquity} \quad \delta\epsilon & 0.231 & 0.223 \\
\text{Nutations in longitude} \quad \delta \psi & 0.383 & 0.253 \\
\text{Large liquid core (FCN=242d)} & 0.01 & 0.008 \\
\text{Small liquid core (FCN=285d)} & 0.005 & 0.004 \\
\text{Precession } (\Delta\dot\psi = 2  \text{ mas/y}) & 0.051 & 0.051 \\
\text{LOD variations} \quad \delta \phi & 0.783 & 0.574 \\
\text{Polar motion} & 0.006 & 0.024 \\
\hline
\end{array}$
\caption{Maximal value of the MOP signature in the Doppler observable (in mm/s) for 
the InSight and ExoMars missions. For each mission, the nominal mission time interval is used. 
The lander position is in Elysium Planitia for InSight and in Oxia Planum for ExoMars.}
\label{tablsi}
\end{table}

\begin{figure}[!htb]
\centering
\includegraphics[height=8cm,width=13.3cm]{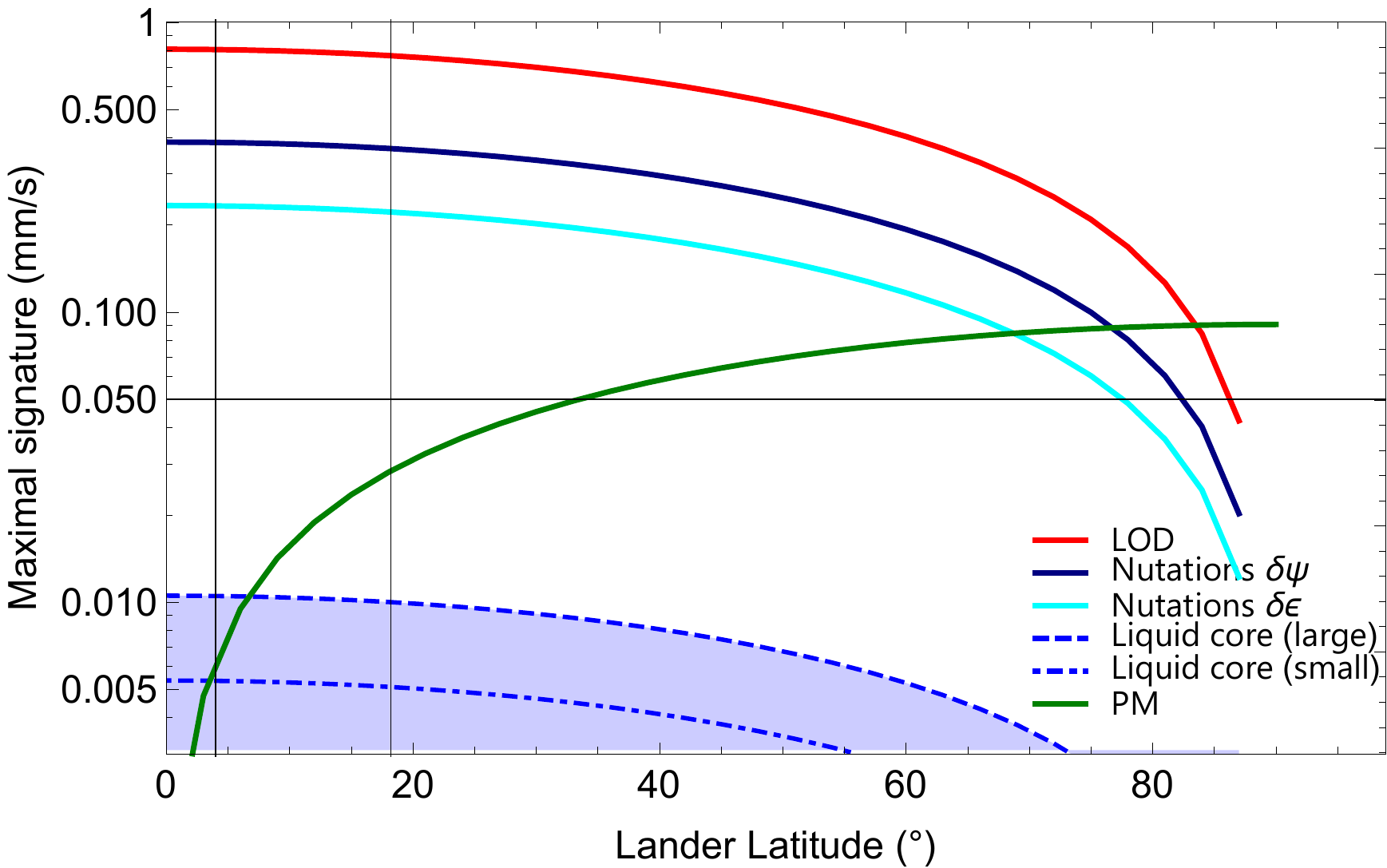} 
\caption{Maximal signature over a very long period of the nutations $\delta\epsilon$ and $\delta\psi$, 
the liquid core effect through nutations,
the LOD variations and the polar motion (PM) in the Doppler observable 
as a function of the lander latitude.
Two different liquid core sizes have been considered.
The two vertical lines indicate the landers latitude 
while the horizontal line shows the typical noise on Doppler measurements ($\sim 0.05$~mm/s).}
\label{fig_sensilat}
\end{figure}

Tab.~\ref{tablsi} shows that the maximal signatures are smaller for ExoMars 
than for InSight, except for the polar motion signature, mostly because the ExoMars nominal mission 
does not cover one full Martian cycle and because the chosen ExoMars 
location is further from the equator than the InSight landing site.
The signature of the precession rate is very close because a small decrease caused by a higher latitude
is compensated by a small increase because the beginning of the ExoMars observation time is later with 
respect to J2000.
The numerical values of Tab.~\ref{tablsi} can be compared to the typical noise on Doppler 
data, which is $\sim 0.05$~mm/s for a Doppler counting interval of 60 seconds.
Fig.~\ref{fig_sensilat} shows that the maximal value of these signatures depends on the lander latitude.
Some rotation parameters like the nutation, the precession and the LOD variations vary with $\cos \theta$
with a larger signature at the equator while the polar motion varies with $\sin \theta$ with a larger 
signature at the pole.
\\

The maximal signature of the nutations in the Doppler signal is up to $0.38$~mm/s, 
coming from the signature of the nutations in longitude $\delta\psi$.
The nutation in obliquity signature is smaller (up to $0.23$~mm/s),
see Fig.~\ref{fig_sensilat}. 
There is a difference of 6 hours between the maxima in the nutation in obliquity and 
in longitude, therefore they cannot be maximal at the same time.
The maximum of the nutation signature is smaller than the product of the maximal value 
of the component of Eqs.~(\ref{eq_signeps}) and (\ref{eq_signpsi})
(i.e.\ $\sin\delta_E^{max} = 0.42$ and the maximum of $\delta\epsilon(t)$ or 
$\delta\psi(t)$)  
because the declination and the nutation do not have their maximum at the same time.
\\

\begin{figure}[!htb]
\centering
\includegraphics[height=6.5cm,width=8.3cm]{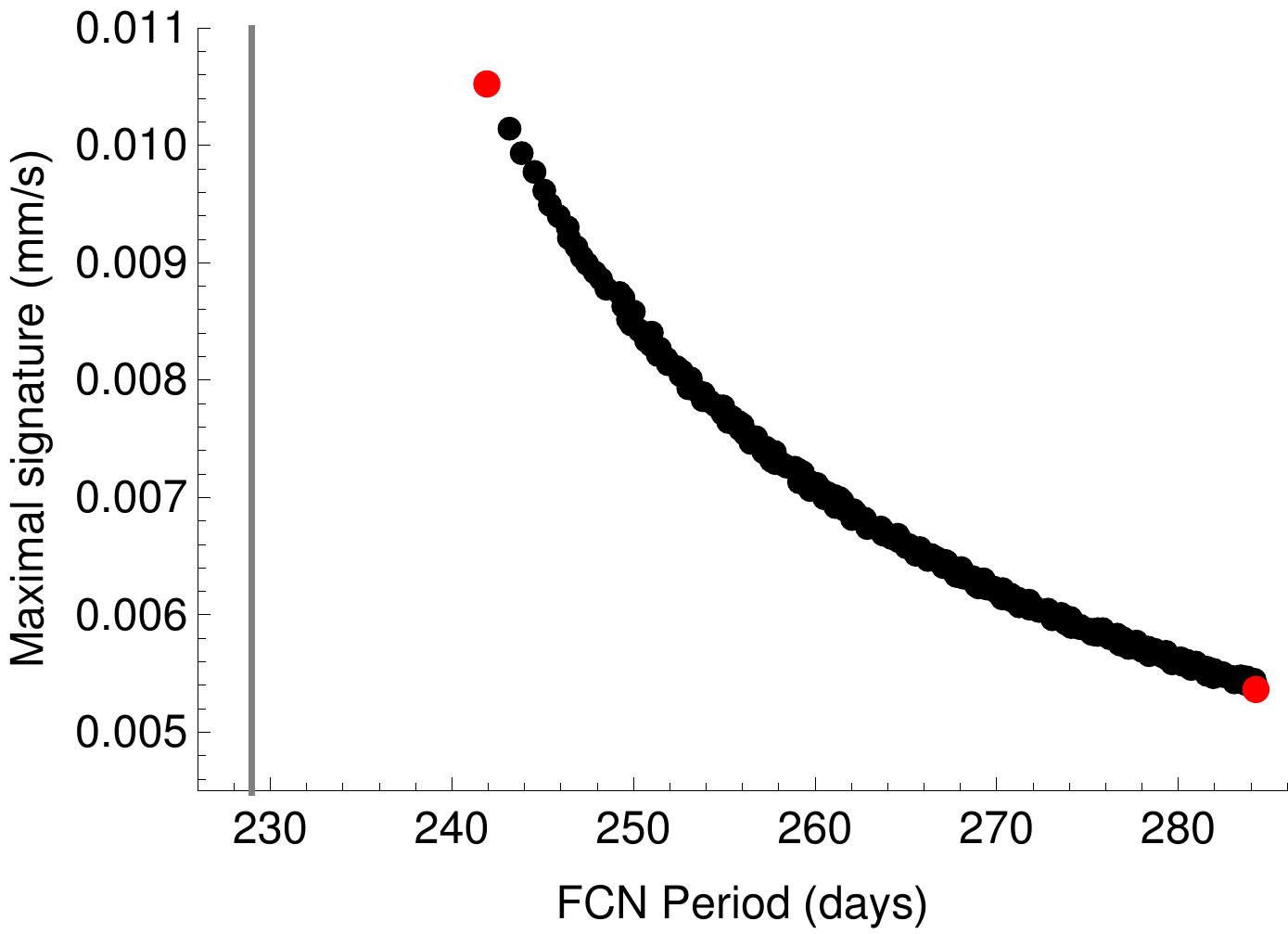}
\caption{Maximal signature of the liquid core in the Doppler observable (in mm/s) 
as a function of the period of the FCN for an equatorial lander. 
The vertical line shows the terannual nutation period while
the red dots show the two models considered in the tables.}
\label{fig_sensinut}
\end{figure}

The liquid core signature in the Doppler observable is much smaller than the nutation signature 
and largely depend on the characteristics of the core because of the 
resonant effects in the transfer function. 
The maximal signature of the liquid core in the Doppler observable is plotted as a function 
of the FCN period on Fig.~\ref{fig_sensinut}.  
The signature is larger (up to 0.01~mm/s) when the FCN is close to the terannual frequency.
This can also be seen on Fig.~\ref{fig_liquid}.

A shift on the precession equal to the present uncertainty on the precession 
rate (2.1 mas/y, corresponding to a period shift of $47$ years, Konopliv et al., 2016) 
changes the Doppler observable by about $0.05$~mm/s after $20$ years.
\\

Whatever the lander latitude $\theta$ below 80$^\circ$, the LOD signature is well above the
other signatures. It culminates at about $0.78$~mm/s for a lander located close to the equator.
This explains why this rotation parameter has already been measured using the previous Martian landers.
\\

Assuming a Chandler wobble amplitude of $1.2$ m,
the maximal signature of the polar motion for a lander latitude of $17^\circ$ is 0.024~mm/s and
$0.006$~mm/s for a lander latitude of $4^\circ$.  
The signature increases as the latitude increases and an equatorial lander is not sensitive 
to the polar motion. 
If the lander is close to the pole, the maximal signature is close to 0.1~mm/s.
\\

A tidal displacement of 1 cm on Mars surface has a very small signature in the
Doppler observable of about $0.0007$~mm/s for an equatorial lander.
Since the radial displacement associated with the largest tidal wave on Mars is on the order of 5 mm 
(Van Hoolst et al., 2003), this is not observable with the present Doppler accuracy.
\\

The factor of proportionality between a parameter (expressed in surface displacement on the planet's surface) 
and its signature in the Doppler observable
is close to $\Omega$ for the LOD variations, the X and Y-coordinates, the tidal displacement
and the polar motion.
This factor is about $\delta_E \, \Omega$ for the nutation in obliquity and 
$\delta_E \, \Omega \, \sin \epsilon$ for the nutation in longitude.
For Mars, the rotation rate is large ($\Omega = 7.088 \, 10^{-5}$ rad/s).
For example if a $1$ meter shift is applied to the X or Y-coordinates, 
the Doppler observable changes by maximum 0.07~mm/s. 
However if the Z-coordinate of the lander changes by 10 meters, the signature in 
the Doppler is negligible (smaller than 0.001~mm/s).
\\

The expressions given before are first order expressions linearized with respect to the MOP. 
The difference between the full and the first order expressions is smaller than 
2\% of the signal.
We also compared these analytical expressions with outputs from the GINS software 
(G\'eod\'esie par Int\'egrations Num\'eriques Simultan\'ees), a numerical software able to 
process planetary geodesy data, and  the differences between the outputs of the two methods
are also small ($< 4\%$).

\subsection{Numerical evaluation of the rotation signature in the Doppler observable 
for a lander on another planet}
We can apply the expressions of the rotational variations signature in the lander-Earth direct 
radio link to
other bodies of the solar system and estimate the order of magnitude of the signature, 
see Tab.~\ref{tab_ss}. 
The lander on the planetary surface has to be equipped with a transponder  
and the communication is direct-to-Earth (DTE). 
The previous signature expressions cannot be applied to Doppler observable between a lander 
and an orbiter, or between an orbiter and the Earth.
The largest rotational variation of bodies in spin-orbit resonance is usually 
the forced libration at the orbital frequency.
This libration amplitude depends on the interior structure and properties and on the presence 
of a liquid layer. 
Additional forced librations with other frequencies also exist.
The obliquity is another parameter important to observe, because if the body is in
the Cassini state, the equilibrium obliquity depends on the body moment of inertia.

\begin{table}[h]
\centering
\begin{tabular}[]{|c|c l|c|}
\hline
Body &  \multicolumn{2}{c|}{Amp. of the main libration} & Signature in the DTE Doppler obs. \\
\hline 
Mercury   & 38.5 as = 0.0107$^\circ$ & = 455 m & 0.56~mm/s \\
Phobos    & 1.1$^\circ$            & = 216 m & 98~mm/s         \\
Europa    & 0.0039$^\circ$         & = 108 m & 2.2~mm/s \\
Ganymede  & 0.00015$^\circ$        & = 7 m   & 0.07~mm/s \\
Callisto  & 0.0002$^\circ$         & = 9 m   & 0.039~mm/s \\
Enceladus & 0.12$^\circ$           & = 528 m & 28~mm/s \\
Titan     & 0.0007$^\circ$         & = 32 m  & 0.15~mm/s \\
\hline 
\end{tabular}
\caption{Amplitude of the libration at the orbital period and its maximal signature in the
Doppler observable between the Earth and a lander on the planetary body. 
The lander is assumed to be on the equator which maximized the libration displacement.
When the libration amplitude is not yet measured, we use estimation coming from a theoretical model.}
\label{tab_ss}
\end{table}

Mercury is in a 3:2 spin orbit resonance and orbits around the Sun with a period of 88 days.
The main libration is the annual libration, with an amplitude of 38.5 as (Margot et al., 2012).
Using Eq.~(\ref{eq_UTsign}), the maximal signature in the Doppler observable of this libration is 0.56~mm/s.
The Earth declination seen from Mercury varies between -11$^\circ$ and 11$^\circ$.
Mercury has an obliquity of 2.04 arcmin (Margot et al., 2012).
Using Eq.~(\ref{eq_signeps}), the maximal signature of the obliquity in the Doppler signal is about 0.33~mm/s,
smaller than the libration signature.
For Mercury, like for Mars, the change of the planet velocity caused by the rotation parameter 
is also the dominant term in the Doppler observable signature. 
However since the distance to Earth is usually smaller for Mercury than for Mars, 
the geometrical effect for Mercury is not as small as for Mars.
\\

Since we evaluate only the change of the lander velocity due to the rotation parameter, 
the previous equations are valid if the lander is far from the receiving antenna on Earth.
If the lander-receiver distance is smaller, the change of the line-of-sight direction must be 
taken into account and increases the Doppler observable signature.
Since this line-of-sight change is not modeled here and since the
geometry of the Earth-Moon configuration is very different from the Earth-planet
configuration due to the synchronous rotation, we do not evaluate the 
signatures for a lunar lander.
\\

If the lander is on Phobos surface, the signature of the libration amplitude is huge 
(about 100~mm/s), because Phobos rotation rate and its libration amplitude are both large.
\\

For the icy moons around Jupiter and Saturn, the velocity effect is larger than 
the geometry effect. 
Each libration signature in the Doppler observable is proportional to the moon radius,
to the diurnal frequency and to the moon libration amplitude. 
The variation of the Earth declination does not largely affect the libration signature.
For Enceladus, the estimated signature is large (about 28~mm/s) because the measured libration
amplitude is large ($0.12^\circ$, Thomas et al., 2016).
For the other moons, the libration amplitudes have not yet been measured but we use order of magnitude 
from theoretical estimations.
The signature in the Doppler observable varies between 0.07 and 2.2~mm/s, see Tab.~\ref{tab_ss}.
These signatures are larger than the typical noise on Doppler data.
\\

A tidal displacement of 60 cm on Europa surface has a signature in the
lander-Earth Doppler observable of about $0.007$~mm/s for an equatorial lander.

\section{MOP signature in the range observable}
\begin{figure}[!htb]
\hspace*{-1cm}
\includegraphics[height=18cm,width=18cm]{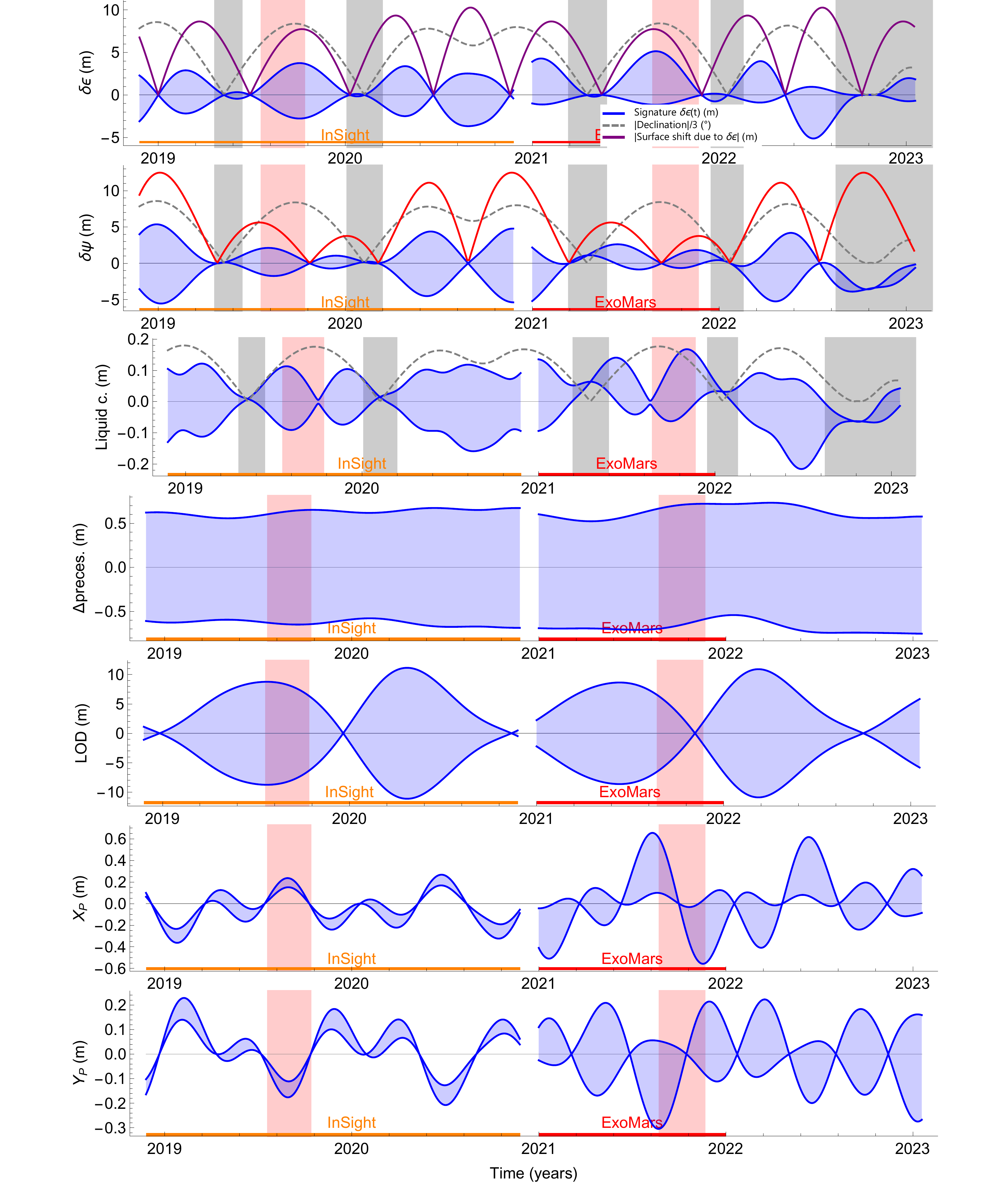} 
\caption{Temporal evolution of the MOP signatures in the range observable (in meters),
respectively: the nutations in obliquity $\delta\epsilon$, 
the nutations in longitude $\delta\psi$,
the liquid core effect in the nutations for a large core, 
a precession rate $\dot \psi$ difference of 2 mas/y, 
the LOD variations and the polar motion ($X_P$ and $Y_P$). 
On the left part of the graph, the InSight lander is in Western Elysium Planitia 
($\theta = 4^\circ$N, $\lambda = 136^\circ$E)
and on the right part, the ExoMars lander is in Oxia Planum 
($\theta = 16.6^\circ$N, $\lambda = 335.45^\circ$E).
In the nutation plots, the gray dashed line shows the Earth declination $\delta_E$.
The red and purple curves are the surface displacements due to the 
nutation in longitude and nutation in obliquity, respectively. }
\label{fig_sensiRAN}
\end{figure}

In this section, we give the expressions of the MOP and lander position signature in the range observable.
The instantaneous range observable is equivalent to the distance between the lander and the antenna 
on Earth at a given time.
The MOP signature in the range observable $\Delta \rho_{MOP}$ is, to the first order in the rotation parameter,
the vector change of position caused by the MOP $\mathbf{\Delta R}$ projected on the Earth-Mars 
line-of-sight $\mathbf{\hat{1}_{EM}}$.
\begin{equation}
\Delta \rho_{MOP} =  \mathbf{\hat{1}_{EM}} \cdot  \mathbf{\Delta R}
\label{eq_ranobs} 
\end{equation}
This expression is independent of the Earth-Mars distance.
The variables that appear in the following expressions are the position of the Earth with respect to Mars 
(i.e.\ the Earth declination $\delta_E$, the right ascension $\alpha_E$ and the hour angle $H_E$) 
and the lander position.

Since the Doppler observable can be seen as the time derivative of the range observable, 
the parameter signature in the Doppler observable is the time derivative of the parameter signature 
in the range observable. 
Therefore the global shapes of the signature of these two observables are very similar, 
this can also be seen on Figs.~\ref{fig_sensi} and \ref{fig_sensiRAN}.
The range signatures may have additional shift or long-term variations that disappear in the Doppler 
observable signature.
Again the MOP signatures in the range observable have a diurnal variation and
are proportional to the MOP themselves.\\
The Same Beam Interferometry (SBI) observable is described in appendix \ref{sec_SBI}. This corresponds
to a difference of two range observables.

\subsection{The signature of the nutations in the range observable}
The signatures of the nutation in longitude $\delta\psi$ and nutation in obliquity 
$\delta\epsilon$ in the range observable are
\begin{eqnarray}
\Delta \rho _{\delta\epsilon} & =
& R \, \delta\epsilon \,(\cos \delta_E \, \sin \alpha_E \, \sin\theta 
- \sin \delta_E \, \cos \theta \sin(H_E + \alpha_E)\,) 
\label{eq_signRANeps}\\
\Delta \rho_{\delta\psi}& =
& R \, \delta\psi\, \sin \epsilon \,( - \cos \delta_E \, \cos \alpha_E \, \sin\theta 
+ \sin \delta_E \, \cos \theta \cos(H_E + \alpha_E)\,)
\label{eq_signRANpsi}
\end{eqnarray}
The first term of these expressions changes slowly and is null if the lander is close to the equator 
while the second term has a diurnal variation and is dominant.
Close to the equator, the dependence of the range signature on the declination is the same 
as the Doppler signature:
when the Earth declination is null, the nutation signature in the range observable is null too.

The liquid core signature in the range observable is the difference between the range observables 
with and without the nutation transfer function, each part being the sum of the contributions of the 
nutation in longitude and in obliquity (Eqs.~(\ref{eq_signRANeps}) and (\ref{eq_signRANpsi})).

\subsection{The signature of the precession in the range observable}
The signature of the precession rate in the range observable is
\begin{equation}
\Delta \rho_{\dot\psi} = R \, \dot\psi \, t \,( 
 \cos \delta_E \, \sin H_E \, \cos\theta \, \cos\epsilon 
- \cos \delta_E \, \cos \alpha_E \, \sin\theta \, \sin \epsilon
+ \sin \delta_E \, \cos \theta \cos( H_E + \alpha_E)\, \sin \epsilon \,)
\label{eq_signRANpsip} 
\end{equation}
The precession signature increases with time but on an interval like the mission duration 
which is short with respect to the precession period of about $170\,000$ years, 
this increase is relatively small. 
The diurnal variations dominate the signature.
The second and third terms are equivalent to the signature of the nutation in longitude $\delta\psi$.
The additional term here with respect to the $\delta\psi$ signature in the range observable
(the first one) depends on $\cos\epsilon$ while 
the others depend on $\sin \epsilon$.
The largest term is the first one ($ \cos \delta_E \, \cos\theta \, \cos\epsilon\approx 0.9$), 
then the third one. The second one (without diurnal variations) is negligible.

\subsection{The signature of the LOD variations in the range observable}
The signature of the LOD variations in the range observable is
\begin{equation}
\Delta \rho _{LOD} = R \,\delta\phi \, \cos \theta \,\sin H_E \, \cos\delta_E 
\label{eq_UTRANsign}
\end{equation}
When the Earth culminates in the lander sky ($H_E=0$), the LOD signature is null.
Therefore the range maxima happen 6 hours later than the Doppler maxima.

\subsection{The signature of the lander position in the range observable}
The signatures of the lander coordinates in the range observable are
\begin{eqnarray}
\Delta \rho _{XYZ} & = & 
-  \, X \, \cos\delta_E \, \cos(\phi - \alpha_E)  
+  \, Y \, \cos\delta_E \, \sin(\phi - \alpha_E) 
-  \, Z \, \sin \delta_E
\label{eq_posRANsign}
\end{eqnarray}
The signatures of the X and Y-coordinates have a diurnal variation.
The signature of the Z-coordinate in the range observable, contrary to the Doppler signature, is not null
and has no diurnal variation. It has long term variations because of the $\sin \delta_E$.
Therefore some range measurements when the Earth declination is not null can be useful to constrain 
the Z-coordinate of a lander.

\subsection{The signature of the polar motion in the range observable}
The signatures of the polar motion in the range observable are
\begin{eqnarray}
\Delta \rho _{X_P} & = & 
R \, X_P \, (- \sin\delta_E \cos\theta \cos\lambda + \cos\delta_E \sin\theta \cos(\alpha_E - \phi) )
\label{eq_PMRANsign}\\
& = & R \, X_P \, (- \sin \delta_E \cos\theta \cos\lambda + \cos\delta_E \sin\theta \cos(\lambda -H_E) )\\
\Delta \rho _{Y_P} & = & 
R \, Y_P\, (\sin \delta_E \cos\theta \sin\lambda - \cos\delta_E \sin\theta \sin(\alpha_E - \phi) )
\label{eq_PMRANsign2}
\end{eqnarray}
The first part of these expressions ($\sin\delta_E \cos\theta \sin\lambda$ or 
$\sin\delta_E \cos\theta \cos\lambda$) is large if the lander is close to the equator (up to 0.3 m)
while the second term has a diurnal variation and is much smaller for an equatorial lander.

\subsection{Comparison to the literature}
Estefan et al. (1990) gave a first order analytical expression of the range observable (see their equation 18).
The coordinates they use to describe the position of the line-of-sight are Earth based,
and not Mars based, which give more lengthy expressions. 
However it is possible to transform their expressions to ours by converting 
their angles ($\delta$, $\delta_0$, $\alpha-\alpha_0$,
$W + \lambda$) to our Mars equatorial coordinates ($\delta_E$, $\alpha_E$, $H_E$) 
using spherical trigonometry relations in the spherical triangle.
\\
They do not consider the MOP signatures. 
However they compute the derivatives of the range observable with respect to the rotation rate, 
the pole orientation, polar motion (actually their definition of polar motion is 
different than ours), lander position and ephemeris.
If expressed with the same angles, the lander position derivatives of Estefan et al. (1990)
is similar to our Eqs.~(\ref{eq_posRANsign}).
Since the rotation angle and the longitude are computed in the same plane but with a different 
reference point, their derivative of the range observable with respect to the rotation angle is 
the same as our derivative with respect to the lander longitude.

\subsection{Numerical evaluation of the signatures in the range observable for a Martian lander}
The maximal values of the MOP signatures in the range observable are given in 
Tab.~\ref{tablsiRAN} for the InSight and ExoMars missions. 
They can also be viewed on Fig.~\ref{fig_sensiRAN}.
\begin{table}[!htb]
\centering
$\begin{array}{|c|c|c|}
\hline
\text{MOP signature in the range observable} & \text{InSight} & \text{ExoMars} \\
\hline
\text{Nutations in obliquity } \quad \delta\epsilon & 3.7 \textrm{ m}  & 5.1 \textrm{ m} \\
\text{Nutations in longitude } \quad \delta\psi     & 5.5 \textrm{ m}  & 5.2 \textrm{ m} \\
\text{Large liquid core}                            & 0.16 \textrm{ m} & 0.17 \textrm{ m} \\
\text{Small liquid core}                            & 0.08 \textrm{ m} & 0.11 \textrm{ m} \\
\text{Precession ($\Delta\dot\psi =$ 2 mas/y)}      & 0.69 \textrm{ m} & 0.72 \textrm{ m} \\
\text{LOD variations} \quad \delta\phi              & 11.1 \textrm{ m} & 8.3 \textrm{ m} \\
\text{Polar motion ($X_P$ and $Y_P$ together)  }    & 0.36 \textrm{ m} & 0.65 \textrm{ m} \\
\hline
\end{array}$
\caption{Maximal values of the MOP signatures in the range observable (in meters) 
for the InSight and ExoMars missions. For each mission, the nominal mission time 
coverage is used.}
\label{tablsiRAN}
\end{table}

As for the Doppler signature, the parameter that has the largest signature in the range observable 
is the LOD variations with a signature of about 11 m.
Then the nutation has a maximal signature of about 3.7 m to 5.5 m.
The nutation signature in the range observable is smaller than the nutation amplitude
and the lander displacement given in Tab.~\ref{tab2} because of the reduction proportional to 
the sinus of the Earth declination.
The liquid core signature is much smaller, about 0.17 m.
The polar motion signature is larger than the liquid core signature (up to 0.7 m) and 
has a long term modulation while its diurnal variations is smaller. 
This is highly dependent on the lander position.
If the precession rate shifts by 2.1 mas/y,  
the precession signature in the range observable is max 0.74~m around the year 2020. 
\\

The ratio between the maximal signature of a MOP in the range observable and its signature 
in the Doppler observable is $1/\Omega$. This assumes that the MOP only has a long term behavior,
like seasonal variations. 
Therefore, as explained in Cical{\`o} et al. (2016), if the ratio of their noise level 
$\sigma_{range}/\sigma_{Doppler}$ is smaller than $\sim 1.4\, 10^{-4}$ sec, 
then range measurements are more sensitive to the MOP, while if this ratio is larger 
than $\sim 1.4\, 10^{-4}$ sec, then the Doppler measurements have to be privileged.

\subsection{Numerical evaluation of the signatures in the range observable for other bodies}
Using Eq.~(\ref{eq_UTRANsign}) and similarly as the Doppler observable study, we can numerically evaluate
the signature in the range observable of a libration at the orbital period
for a lander on another planet or moon of the solar system.
The typical noise on range data between the Earth and a lander on another body 
is about a few meters.
The maximal value of the rotation parameter signature is simply the maximal displacement 
on the body surface caused by the libration.
If we use the order of magnitude of the libration amplitude given in Tab.~\ref{tab_ss},
the maximal signature in the range observable for an equatorial lander is given by the second column.
It varies from a few meters up to 528 meters for Enceladus.
The signature for Mercury is also large, 455 m.
The maximal signature of Mercury obliquity in the range observable is about 270 m,
smaller than the libration signature.
All the signatures are modulated by the diurnal rotation and by variations 
at long period due to the configuration geometry.

\section{Discussion}

In the previous sections, we present the analytical expressions for the MOP signatures.
These analytical expressions are useful for gaining general insight into the behavior 
of the signatures in the data. 
We can anticipate when each signature is null or maximal.
When accumulating data and assuming no correlations between parameters, 
each additional observation time with a non-null derivative add information to the system, 
and its a posteriori uncertainty will therefore decrease.
If an observation time with no signature is added to the series (for example for the nutation, 
a period where the declination is null), then 
the a posteriori uncertainty for this rotation parameter will not decrease at this time.
\\
The MOP derivatives of the observable, useful for the construction of the information matrix, 
are straightforward to compute from the MOP signatures expressions because the latter are developed 
up to the first order in the MOP.
\\
The expressions for the signatures can also be used 
to compare the sensitivity of different observables (Doppler versus range versus SBI observables), 
to understand the role of various configuration parameters, 
and to evaluate the rotation signature on other bodies of our solar system using geodesy.

Defining the best observational schedule is useful for the mission planning of 
the geodesy experiments. 
On a long time interval, 
an observation strategy that favors observation times that maximize the signatures 
may be inefficient for the MOP estimation because of large correlations between the MOP signatures.
A large signature of a rotation parameter in the Doppler observable cannot be automatically associated
with a small uncertainty in the estimation of this parameter. 
In order to estimate the level of precision that can be obtained for each MOP, 
a full numerical simulation with the resolution of all the MOP together is needed 
(see for example Le Maistre et al., 2012). 
In these simulations, noisy measurements are simulated according to an observation schedule
depending on multiple observational and technical constraints, using a dedicated software.
The outputs of these global simulations are the a posteriori uncertainties on each parameter 
and their correlations.
These simulations are extensive because the results will depend 
on the considered mission operational parameters like the 
starting date of the mission, the mission duration, the blackout periods, 
the availability of the ground stations, the timing of observations, 
the tracking pass length and occurrence, 
the assumption for the noise behavior, the chosen elevation and azimuth in the
lander sky, etc.
This is currently being done in the frame of the preparation of two geodesy experiments: 
InSight and ExoMars.
\\
Nevertheless, knowing the signatures is useful to interpret the results of different 
simulation strategies and to select the best numerical simulations to implement because 
covering a large range for each parameter will be too time consuming.
This study also gives us clues why a new strategy increases or decrease the science return. %case/
\\

We can also analytically investigate the correlations between two different parameters.
If two rotation parameters have exactly the same diurnal and long term signatures, 
it is very difficult to separate them and their correlation is close to $1$ 
even if a lot of data are accumulated.
This is for example the case for the polar motion and the rotation rate variations that have 
the same dependency in the hour angle.
In order to illustrate that, we consider a simple model with only two parameters corresponding to the same frequency 
for the polar motion $X_P$ and for the LOD $\delta \phi$.
As the lander is assumed to be tracked continuously from the Earth hour angle $H_1$ to $H_2$, 
we replace the sum over the observations during one tracking pass by an integration over $H$.
We also assume that the MOP and other parameters like the declination or the right ascension 
do not change on this short timescale.

As a result, each element of the variance-covariance matrix is a function of the same sum depending on 
the long term schedule of observation.
The correlation between these two parameters is independent of this timing factor and is equal to:
\begin{equation}
\frac{ \cos(2 H_1 - \lambda) - \cos(2 H_2 - \lambda) + 2 (H_1 - H_2) \sin\lambda}
{\sqrt{2 (H_1 - H_2 + \cos H_1 \sin H_1 - \cos H_2 \sin H_2) \, (2 H_1 - 2 H_2 - 
\sin(2 H_1 - 2 \lambda) + \sin(2 H_2 - 2 \lambda))}}
\label{eqcorr}
\end{equation}
We compute the information matrix using the MOP derivatives integrated 
from $H_1$ to $H_2$. The variance-covariance matrix is the inverse of this 2x2 matrix. 
The correlation between the two parameters is extracted from the covariance matrix, 
it is equal to the covariance (the out-of-diagonal element) divided by the square root of 
the variances (the diagonal elements). \\
The correlation given by Eq.~(\ref{eqcorr}) is close to $1$ in the limit of a very short hour angle range.
Usually observation sessions last around 1 hour or less, which is shorter than one day.
Therefore observing everyday at the same solar time induces high correlations between parameters.
A better strategy would be to have observations that correspond to different solar times.
Increasing the hour angle interval also decreases the correlation.
Since the correlation for this particular pair of parameters is independent of time, 
the correlation is the same if 10 or 1000 passes are added, assuming 
the same hour angle range is chosen for all the passes. If the hour angle range changes 
from one pass to the other, no general expression can be found, a numerical study
has be done for each particular timing strategy.
This suggests that in order to enrich the Doppler geometry and avoid repeatability of observations,
it is better to favor various line-of-sight directions during the tracking sessions.
\\

\begin{figure}[!htb]
\includegraphics[height=6cm]{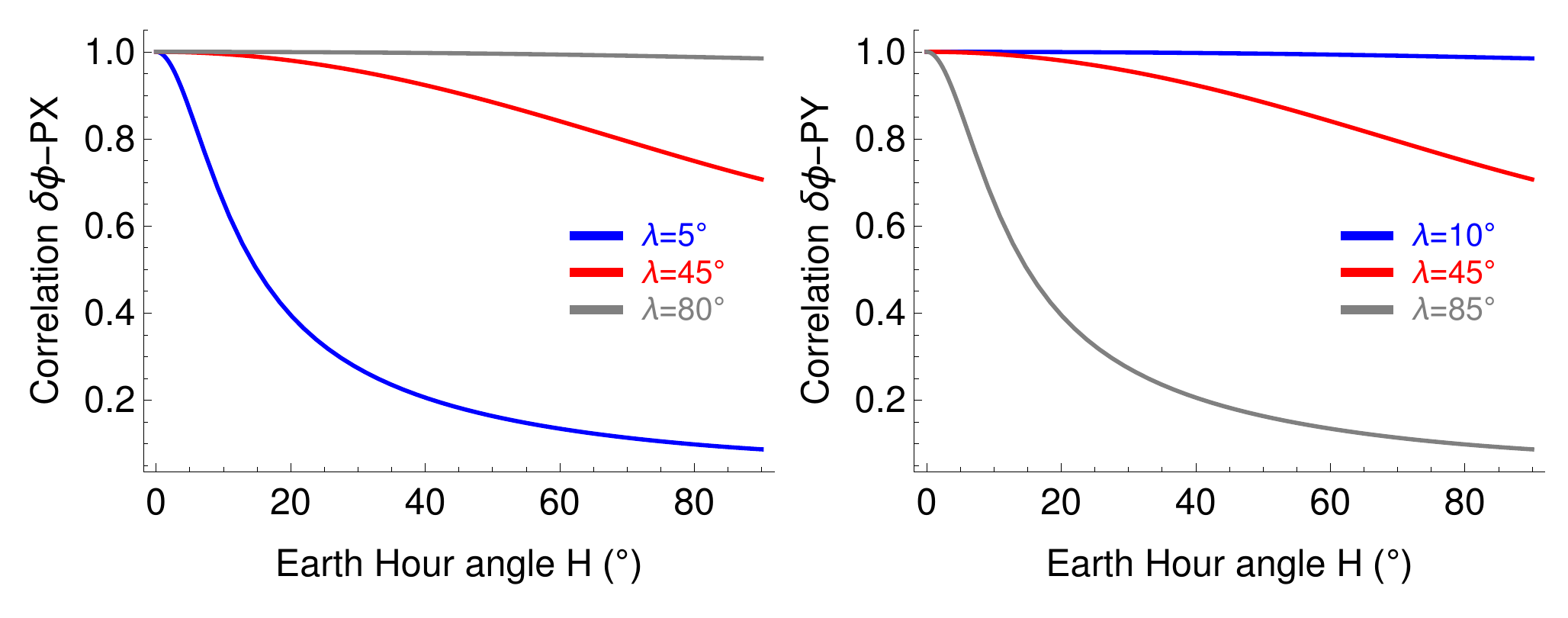} 
\caption{Correlations between the LOD and the polar motion $X_P$ (left) and $Y_P$ (right)
as a function of the Earth hour angle $H$ of the pass around the upper culmination. 
The pass is assumed to be symmetric around the upper culmination of the Earth (from $-H$ to $H$).
Three different lander longitudes have been compared.}
\label{fig_corr}
\end{figure}

If the observation interval is symmetric around the upper culmination of the Earth in the 
lander's sky (from $-H$ to $H$), the correlation between $X_P$ and $\delta \phi$ for the 
same frequency is 
\begin{equation}
\sin \lambda \, \sqrt{\frac{2 H + \sin(2 H)}{ 2H - \sin(2 H) \cos(2 \lambda)}}
\end{equation}
while the correlation between $Y_P$ and $\delta \phi$ is 
\begin{equation}
\cos \lambda \, \sqrt{\frac{2 H+\sin (2 H) }{2H + \sin (2 H) \cos (2\lambda)}}
\end{equation}
see Fig.~\ref{fig_corr}.
Therefore since the lander longitude $\lambda$ is fixed, solving for both coordinates of the 
polar motion when only one lander is on Mars surface may lead to very high correlations 
between the rotation parameters.
\\

We now consider two nutation amplitudes $\delta \psi_i$ and $\delta \epsilon_i$ corresponding 
to the same frequency, and an hour angle interval from $-H$ to $H$ around the upper culmination. 
We can analytically find the correlation between these two parameters because they have 
the same diurnal behavior.
Assuming that the right ascension is not changing quickly and using the same method as before 
(computation of the inverse of the 2x2 normal matrix), the correlation over a few weeks or less is 
\begin{equation}
\frac{ \sin (2 \alpha_E) \, \sin(2 H)}
{ \sqrt{4H^2 - \cos(2 \alpha_E)^2 \, \sin(2 H)^2}}
\label{eq_cor_nut}
\end{equation}
Again, this expression is close to $1$ in the limit of a very short pass duration.
The correlation decreases to $0$ as we integrate over a longer pass.
Assuming a continuous tracking from Earth rise to Earth set, 
then the correlation between $\delta \psi_i$ and $\delta \epsilon_i$ would be close to $0$.
\\

When the temporal dependence of two parameters are different, for example for the polar motion
and the nutation, no general behavior of the correlation can be anticipated 
with a simple expression like Eqs.~(\ref{eqcorr}) to (\ref{eq_cor_nut}).

If more than two parameters are considered, then the equation for the correlation is not 
as easy to express 
as before and a full numerical evaluation is needed, depending on the chosen 
observation schedule and the mission operational parameters.

\section{Conclusion}
Using the analytical expressions of the Mars Orientation Parameters (MOP) signatures 
in the Doppler observable, we can anticipate the moments where the MOP signatures are 
maximal, on a daily basis and on longer interval.
The MOP signatures in the Doppler and the range observable are proportional to the MOP signal. 
Therefore, the periodicities of the MOP (mostly annual, semiannual, terannual, 
the Chandler period of about 200 days for the polar motion etc.) can also be directly seen 
in the signatures.
\\

Some signatures like the nutations depend on the Earth declination.
The signature is larger if the Earth is outside the equatorial plane of Mars.
The nutations signature is maximized when the Earth declination is large. 
Additionally, the noise on the Doppler measurement caused by the plasma is smaller 
when the Sun-Earth-Probe angle is large.
The combination of these two constraints in order to have a larger signal-to-noise ratio for the 
nutation amplitude corresponds to the beginning (2018.9-2019.25) and end (2020.3-2021) of the InSight mission.
For the ExoMars mission, this corresponds to approximately 2021-2021.1 and 2021.5-2021.8.
\\

In every signature, there is also a diurnal modulation since the main motion 
affecting the lander is the diurnal rotation of Mars. 
Therefore every day, each MOP signature has a maximum and a minimum signature. 
The daily maxima in the signatures of the nutations in obliquity and in longitude 
are separated by 6 hours, therefore it is impossible to maximize both at the same time.
The daily maxima in the length-of-day variations signature in the Doppler observable 
happen when the Earth is at its highest point in the lander sky.
\\

For an equatorial lander, the motions with the largest signature in the Doppler observable are 
caused by the length-of-day variations, the precession rate and the rigid nutations (smaller than 0.8 mm/s).
The polar motion and the liquid core signatures have a much smaller amplitude (up to 0.01 mm/s).
If the lander latitude increases, the polar motion signature is enhanced while 
the other signatures decrease.

\section*{Acknowledgments}
We thank Attilio Rivoldini for the computation of the transfer function and Rose-Marie Baland
and Sebastien Le Maistre for the helpful comments.
We also thank two anonymous reviewers for their thoughtful comments and helpful suggestions.
This work was financially supported by the Belgian PRODEX program
managed by the European Space Agency in collaboration with the Belgian
Federal Science Policy Office.

\appendix

\section{ The nutation signature in the Doppler observable: mathematical development}
\label{ap1}
Assuming there is no LOD variations and no polar motion, we describe here the method to find 
the expression of the nutation signature in the Doppler observable.
However this development can be applied similarly to each MOP.\\

If the vector $\mathbf{r}$ is the fixed lander position in the body fixed reference frame,
the lander position in the inertial frame is $\mathbf{R} = \mathbf{M} . \, \mathbf{r}$. 
Its velocity in the inertial frame is: 
\begin{eqnarray}
\hspace*{-2cm}
\mathbf{V}_{lander} &=& \frac{d\mathbf{M}}{dt} \, . \, \mathbf{r} \\
&\approx& \dot\phi \, R_Z( - \psi) . R_X(-\epsilon) . \frac{d R_Z(-\phi)}{d\phi} . \, \mathbf{r}
\nonumber
\end{eqnarray}
The terms proportional to $\dot\psi$ and $\dot\epsilon$ have been neglected here since they are 
much smaller than $\dot\phi$ ($\approx$ the diurnal rotation $\Omega$). 

To the first order in the nutations in obliquity $\delta\epsilon= \epsilon-\epsilon_0$,
the lander velocity in a non-rotating frame based on the Mars Mean Equator of J2000 (MME2000) is
\begin{eqnarray}
\mathbf{V}^{MME2000} &=& \Omega\; R^T_X(-\epsilon_\mathbf{0}) . R_X( - \epsilon) . \frac{d R_Z(-\phi)}{d\phi} 
. \, \mathbf{r} \\
&=& \Omega\; R\,\cos \theta \left(\begin{array}{c} - \sin(\phi+\lambda)\\
\cos(\phi+\lambda) \\
\cos(\phi+\lambda) \, \delta\epsilon\\
\end{array} \right) 
\end{eqnarray}
The lander velocity $\mathbf{V}^{MME2000}$ is mostly a velocity in the XY plane.
But the lander velocity induced by a nutation in obliquity, in the 
non-rotating frame based on the Mars Mean Equator of 2000, is mainly along the Z axis.
\\

Similarly, the main component of the signature of the nutations in longitude $\delta\psi$ in the 
lander velocity is also the Z component, 
because the dependence on the velocity is only in the Z coordinate. 
\begin{eqnarray}
\mathbf{V}^{MME2000} &=& 
\Omega\; R^T_X(-\epsilon_\mathbf{0}) . R^T_Z( - \psi_\mathbf{0}) . R_Z( - \psi) . 
R_X( - \epsilon_\mathbf{0}) . \frac{d R_Z(-\phi)}{d\phi} 
. \, \mathbf{r} \\
 &=& \Omega\; R\,\cos \theta \left(\begin{array}{c} - \sin(\phi+\lambda)\\
\cos(\phi+\lambda) \\
\sin(\phi+\lambda) \,\delta\psi \, \sin \epsilon\\
\end{array} \right)
\end{eqnarray}
This Z component (to the first order in $\delta\psi$) is not as straightforward as for $\delta\epsilon$
because there are $\delta\psi$ dependent terms both in the products of 4 rotation matrices and in the 
$\frac{d}{dt}R_Z( - \Omega \, t + \delta\psi \cos \epsilon)$ matrix, and there is a cancellation between 
some of these terms.\\

In the same inertial frame, the tracking station-lander direction is approximated by the Earth-Mars direction.
\begin{eqnarray}
\mathbf{LOS}&=& -\left(\begin{array}{c} \cos \alpha_E \cos\delta_E\\
\sin \alpha_E \cos\delta_E\\
\sin\delta_E
\end{array} \right)
\end{eqnarray}
$\alpha_E$ is the Earth right ascension.
The MOP signature in the Doppler observable is the scalar product between the velocity difference
$\mathbf{\Delta V}^{MME2000}$ 
= ($0, \, 0, \, \Omega\, R\,\cos \theta \cos(\phi+\lambda) \, \delta\epsilon$ or 
$\Omega\, R\,\cos \theta \sin(\phi+\lambda) \, \delta\psi\, \sin \epsilon$)
and the LOS vector:
\begin{eqnarray}
\Delta q_{\delta\epsilon} & =& - \,\sin\delta_E \, \Omega\, R\,\cos \theta \cos(\phi+\lambda) 
\, \delta\epsilon
= - \, \sin\delta_E \, \Omega\, R\,\cos \theta \cos(H_E+\alpha_E) \, \delta\epsilon
\label{eq_signeps2}
\\
\Delta q_{\delta\psi}& =& - \sin\delta_E \, \Omega\, R\,\cos \theta \sin(\phi+\lambda) 
\, \delta\psi\, \sin \epsilon
= - \sin\delta_E \, \Omega\, R\,\cos \theta \sin(H_E+\alpha_E) \, \delta\psi\, \sin \epsilon
\label{eq_signpsi2}
\end{eqnarray}
which are equal to Eqs.~(\ref{eq_signeps}) and (\ref{eq_signpsi}).

\section{MOP signature in the Same Beam Interferometry (SBI) observable}
\label{sec_SBI}
The Same Beam Interferometry (SBI) technique, also known as the Inverse VLBI, is based
on the simultaneous tracking of two Martian landers with one large antenna on Earth,
for example a Deep Space Network antenna, and measuring the delay between the two radio signals.
The advantage of the SBI technique is that the two signals cross almost the same media 
at the same time, and therefore taking the difference cancels a large part of the 
common errors, like the Earth atmospheric and ionospheric errors or the plasma delays.
The SBI observable corresponds to the difference of the two ranges between the antenna and 
each lander.
Therefore the rotation parameter signature in the SBI observable is the difference between 
the rotation signatures in the range observables of the two landers.
This technique has never been used up to now. 
It is expected to have an error budget up to 0.7 cm for Martian landers (Gregnanin et al., 2014).

The MOP signature in the SBI observable depends on the latitude and longitude of each lander.
We can replace the local angle $H_E$ by $\phi +\lambda-\alpha_E$ in the previous expressions 
and substitute the latitude and longitude of the lander by their rectangular 
coordinate $(X,Y,Z)$.
For example, the LOD signature in the SBI observable is
\begin{equation}
\Delta SBI_{LOD} = \delta\phi \, \cos\delta_E \, 
(\Delta X \,\sin(\phi-\alpha_E) + \Delta Y \,\cos(\phi-\alpha_E) )  
\label{eq_UTSBIsign}
\end{equation}
where $\Delta X$ and $\Delta Y$ are the difference of X and Y-coordinates of the two landers.
The $\Delta Z$ contribution does not appear in this signature.

The nutation signature in the SBI observable is
\begin{eqnarray}
\Delta SBI _{\delta\epsilon} & =
& \delta\epsilon \,(
- \Delta X \, \sin \delta_E \, \sin\phi
- \Delta Y \, \sin \delta_E \, \cos\phi
+ \Delta Z \, \cos \delta_E \, \sin \alpha_E  \,) 
\label{eq_signSBIeps}\\
\Delta SBI _{\delta\psi}& =
& \delta\psi\, \sin \epsilon \,( 
  \Delta X \, \sin \delta_E \, \cos\phi
- \Delta Y \, \sin \delta_E \, \sin\phi
- \Delta Z \, \cos \delta_E \, \cos \alpha_E  \,) 
\label{eq_signSBIpsi}
\end{eqnarray}
Eqs.~(\ref{eq_signSBIeps}) and (\ref{eq_signSBIpsi}) show that the parts proportional 
to $\Delta X$ and $\Delta Y$ 
are affected by a diurnal modulation and have a negligible contribution if the Earth declination is small.
The $\Delta Z$ contribution has long term variations.

These expressions are similar to the SBI sensitivity given in Kawano et al.~(1999, Eq.~28).
Their variables describing the Earth position with respect to Mars are not equivalent to ours 
(their model is simpler) but it is possible to see the same modulations (the diurnal 
and the long term) by comparing the two methods.

The numerical values of the MOP signatures in the SBI observable 
depend of course on the separation vector between the two landers.
But a general conclusion is that the signature is larger if the distance between the two 
landers is larger too.

\end{document}